\documentclass{aa}
\usepackage{graphicx}

\begin{document}
%

   \title{Luminosity-dependent evolution of soft X--ray selected AGN}

   \subtitle{New {\em Chandra} and {\em XMM--Newton} surveys}

   \author{G\"unther Hasinger\inst{1}
          \and
          Takamitsu Miyaji\inst{2}
          \and
          Maarten Schmidt\inst{3}
          }

   \offprints{G. Hasinger}

   \institute{Max-Planck-Institut f\"ur extraterrestrische Physik,
              Postf. 1312, D-84571 Garching, Germany\\
              \email{ghasinger@mpe.mpg.de}
         \and
             Department of Physics, Carnegie Mellon University, 
             5000 Forbes Avenue, Pittsburg, PA 15213, U.S.A. 
               \email{miyaji@cmu.edu}
        \and
             California Institute of Technology, Pasadena, CA 91125, U.S.A.
              \email{mxs@astro.caltech.edu}
             }

   \date{Received ; accepted }

   \authorrunning{Hasinger et al.}
   \titlerunning{Luminosity-dependent evolution of soft X-ray AGN}

   \abstract{We present new results on the cosmological evolution of
             unabsorbed (type--1) active galactic nuclei (AGN) selected
             in the soft (0.5--2 keV) X--ray band. From a variety of 
             {\em ROSAT}, {\em XMM--Newton} and {\em Chandra} surveys
             we selected a total of $\sim1000$ AGN with an unprecedented 
             spectroscopic and photometric optical/NIR identification
             completeness. For the first time we are able to derive reliable
             space densities for 
             low--luminosity (Seyfert--type) X--ray sources at cosmological
             redshifts. The evolutionary behaviour of AGN shows a strong
             dependence on X--ray luminosity: while the  
             space density of high--luminosity AGN reaches a peak around 
             $z\sim 2$, similar to that of optically selected QSO,
             the space density of low--luminosity AGNs peaks
             at redshifts below $z=1$.
             This confirms previous {\em ROSAT} findings of a 
             luminosity-dependent density evolution. Using a rigorous 
             treatment of the optical identification completeness we 
             are able to show that the space density of AGN with X--ray
             luminosities $L_{\rm x} < 10^{45}$ erg s$^{-1}$ declines significantly
             towards high redshifts.

   }

   \maketitle
%

\section{Introduction}

In recent years the bulk of the extragalactic X--ray background in the 0.1-10
keV band has been resolved into discrete sources with the deepest 
{\em ROSAT}, {\em Chandra} and {\em XMM--Newton} observations (Hasinger et 
al. \cite{has98}, Mushotzky et al. \cite{mus00}, Giacconi et al. \cite{gia01,gia02}, 
Hasinger et al. \cite{has01}, Alexander et al. \cite{ale03}, 
Worsley et al. \cite{wor04}). Optical identification programmes with 
Keck (Schmidt et al. \cite{schm98}, Lehmann et al. \cite{leh01}, 
Barger et al. \cite{bar01,bar03}) and VLT (Szokoly et al. \cite{szo04}, 
Fiore et al. \cite{fio03}) find predominantly
unobscured AGN--1 at X--ray fluxes $S_X>10^{-14}$ erg cm$^{-2}$ s$^{-1}$, and a
mixture of unobscured AGN--1 and obscured AGN--2 at fluxes $10^{-14}>S_X>10^{-15.5}$
erg cm$^{-2}$ s$^{-1}$ with ever fainter and redder optical counterparts, while
at even lower X--ray fluxes a new population of star forming galaxies emerges
(Hornschemeier et al. \cite{hor00}, Rosati et al. \cite{ros02}, 
Norman et al. \cite{nor04}). At optical magnitudes R$>$24 these surveys suffer 
from large spectroscopic incompleteness, but deep optical/NIR photometry can improve the identification completeness significantly, even for the faintest optical counterparts (Zheng et al. \cite{zhe04}, Mainieri et al. \cite{mai05}).

The AGN/QSO luminosity function and its evolution with 
cosmic time are key observational quantities for understanding
the origin of and accretion history onto supermassive black holes,
which are now believed to occupy the centers of most galaxies.
X--ray surveys are practically the most efficient means of finding 
active galactic nuclei (AGNs) over a wide range of luminosity and 
redshift. Enormous efforts have been made by several groups to follow 
up X--ray sources with major optical telescopes around the globe, so
that now we have fairly complete samples of X--ray selected AGNs.
In this work we concentrate on unabsorbed (type--1) AGN selected in
the soft (0.5--2 keV) X--ray band, where due to the previous {\em ROSAT} work
(see Miyaji et al., \cite{paper1,paper2}, hereafter Paper I and II) complete samples exist, with
sensitivity limits varying over five orders of magnitude in flux, and 
survey solid angles ranging from the whole high galactic latitude sky to 
the deepest pencil-beam fields. These samples enable us to construct 
and probe luminosity functions over cosmological timescales, with an
unprecedented accuracy and parameter space.

Conceptually, space densities and luminosity functions are simply 
derived by dividing the observed number of objects $N$ by the volume 
$V$, in which they have been surveyed. The binned luminosity function 
derived in this fashion generally does not represent the center of the bin. 
In paper II, we introduced the estimator $N_{\rm obs}/N_{\rm mdl}$, where
$N_{\rm mdl}$ is the number of objects expected from an analytical 
representation of the luminosity function. Scaling the model value of 
the analytical function at the bin center by the estimator removes the 
binning bias to first order. This method is applied in Sect. \ref{sec:exlf}. 
A quite different method based on $1/V_{\rm max}$ values
(Schmidt \cite{schmidt68}) for individual objects is used in 
Sect. \ref{sec:mxs}.   
It involves a derivation of the zero-redshift luminosity function that 
is free of binning bias. The luminosity function at higher redshifts is 
again derived by employing an analytical representation of the luminosity
function and scaling it by the ratio of observed over expected numbers.
The use of individual $1/V_{\rm max}$ values in this case allows 
accounting for an effective optical magnitude limit beyond which redshifts
have generally not been obtained in some surveys. 

Throughout this work we use a Hubble constant $H_0=70 h_{70}$ 
${\rm km\,s^{-1}\,Mpc^{-1}}$ and cosmological parameters
$(\Omega_{\rm m},\Omega_\Lambda) =$ (0.3, 0.7) consistent with
the WMAP cosmology (Spergel et al. \cite{spe03}).

\begin{table*}[ht]
\caption[]{The soft X--ray sample}
\begin{tabular}{@{}lccccccc@{}}
\hline
Survey$^{\rm a}$ & Solid Angle &  $S_{\rm X14,lim}$ & $N_{\rm tot}$ & $N_{\rm AGN-1}$$^{\rm b}$ & $N_{\rm unid}^{\rm c}$ \\
{}     & [deg$^2$] &[cgs]          &               &                &               &\\
\hline
RBS    & 20391     & $\approx 250$ &   901    & 203  &     0  \\
SA--N   & 684.0--36.0  & 47.4--13.0   &   380    & 134  &     5  \\    
NEPS   & 80.8--70.5   & 12.4--10.1   &   252    & 101  &     1  \\  
RIXOS  & 19.5--15.0   & 10.2--3.0    &   340    & 194  &    14  \\   
RMS    & 0.74--0.32   & 1.0--0.5     &   124    &  84  &     7  \\
RDS/XMM & 0.126--0.087 & 0.38--0.13   &  81    &  48  &     8  \\
CDF--S   & 0.087--0.023   & 0.022--0.0053& 293   & 113  &     1  \\
CDF--N   & 0.048--0.0064  & 0.030--0.0046& 195   &  67  &    21  \\
\hline
Total  &           &               &  2566    & 944  &    57  \\        
\hline

\end{tabular}\label{tab:samp}

$^{\rm a}$ Abbreviations -- RBS: The {\em ROSAT} Bright Survey (Schwope et al. \cite{schw00}); SA--N: {\em ROSAT} Selected Areas North (Appenzeller et al. \cite{app98}); NEPS: {\em ROSAT} North Ecliptic Pole Survey (Gioia et al., \cite{gio03}); RIXOS: {\em ROSAT} International X--ray Optical Survey (Mason et al. \cite{mas00}), RMS: {\em ROSAT} Medium Deep Survey, consisting of deep PSPC pointings at the North Ecliptic Pole (Bower et al. \cite{bow96}), the UK Deep Survey (McHardy et al. \cite{mch98}), the Marano field (Zamorani et al. \cite{zam99}) and the outer parts of the Lockman Hole (Schmidt et al. \cite{schm98}, Lehmann et al. \cite{leh00}); RDS/XMM: {\em ROSAT} Deep 
Survey in the central part of the Lockman Hole, observed with {\em XMM--Newton} (Lehmann et al. \cite{leh01}, Mainieri et al. \cite{mai02}, Fadda et al. \cite{fad02}); CDF--S: The {\em Chandra} Deep Field South (Szokoly et al. \cite{szo04}, Zheng et al. \cite{zhe04}, Mainieri et al. \cite{mai05}); CDF--N: The {\em Chandra} Deep Field North (Barger et al. \cite{bar01, bar03}). 

$^{\rm b}$ Excluding AGNs with $z<0.015$. 

$^{\rm c}$ Objects without redshifts, but hardness ratios consistent with 
type--1 AGN. 

\end{table*}


\section{The X--ray selected AGN--1 sample}

For the derivation of the X--ray luminosity function and cosmological 
evolution of AGN we have chosen well--defined flux--limited 
samples of active galactic nuclei, with flux limits and survey solid 
angles ranging over five and six orders of magnitude, respectively. 
To be able to utilize the massive amount of optical identification work
performed previously on a large number of shallow to deep {\em ROSAT} 
surveys, we restricted the analysis to samples selected in the 0.5--2 
keV band. In addition to the {\em ROSAT} surveys already used in Paper I and II, we included data from the recently published 
{\em ROSAT} North Ecliptic Pole Survey (NEPS, Gioia et al. \cite{gio03}, Mullis et al. \cite{mul04}), 
from an {\em XMM--Newton} observation of the Lockman Hole (Mainieri et al.
\cite{mai02}) and the {\em Chandra} Deep Fields South 
(CDF--S, Szokoly et al. \cite{szo04}, Zheng et al. \cite{zhe04}, 
Mainieri et al. \cite{mai05}) and North 
(CDF--N, Barger et al. \cite{bar01,bar03}). In order to avoid 
systematic uncertainties introduced by the varying and a priori unknown
AGN absorption column densities we selected only unabsorbed
(type--1) AGN, classified by optical and/or X--ray methods.   
We are using here a definition of type--1 AGN, which is largely based on the presence of broad Balmer emission lines and small Balmer decrement in the optical spectrum of the source (optical type--1 AGN, e.g. the ID classes a, b, and partly c in Schmidt et al. \cite{schm98}), which largely overlaps the class of X--ray type--1 AGN defined by their X--ray luminosity and unabsorbed X--ray spectrum (Szokoly et al. \cite{szo04}). However, as Szokoly et al show, at low 
X--ray luminosities and intermediate redshifts the optical AGN classification often breaks down because of the dilution of the AGN excess light by the stars in the host galaxy (see e.g. Moran et al. (\cite{mor02}), so that only an X--ray classification scheme can be utilized. Schmidt et al. (\cite{schm98}) have already introduced the X--ray luminosity in their classification. For the deep XMM--Newton and {\em Chandra} surveys we in addition use the X--ray hardness ratio to discriminate between X--ray type--1 and type--2 AGN, following Szokoly et al. (\cite{szo04}). 

In order to convert the count rates observed in the 0.5--2 keV band 
to unabsorbed 0.5--2 keV fluxes, we assumed a power law AGN spectrum 
with a photon index of $\Gamma=2.0$ and Galactic absorption. Typical 
values of the Galactic neutral hydrogen column density are 
$(0.5-1)\times 10^{20}$ ${\rm cm^{-2}}$ for the deep surveys and 
a maximum of $16\times 10^{20}$ ${\rm cm^{-2}}$ for a small portion 
of the sky covered by the {\em ROSAT} Bright Survey (RBS). Because the
band, in which the 
AGN are selected, is the same as the one for which we calculate the 
fluxes, systematic differences in the true AGN power law indices 
have a negligible effect on the derived fluxes. Assuming spectral 
indices in the range $\Gamma=1-3$, the conversion between the 
observed 0.5--2 keV count rates and the X--ray flux $S_{\rm x}$ (here 
and hereafter, $S_{\rm x}$ represents the 0.5--2 keV flux and 
$S_{\rm X14}$ is the same quantity measured in units of $10^{-14} 
{\rm erg\,s^{-1}\,cm^{-2}}$) varies by less than 10\%.  

The surveys we have used are summarized in Table~\ref{tab:samp}.
A total of 944 X--ray selected type--1 AGN were compiled from 
eight independent samples containing a total of 2566 soft X--ray 
sources. The number of unidentified sources\footnote{We call unidentified
sources those, which do not have a reliable redshift determination,
either through spectroscopy or through photometric redshifts; however
practically all of these have optical or NIR counterparts} in these 
samples is only 86 (of which 57 could be AGN--1), yielding an unprecedented 
identification fraction of 97\%. Due to the extreme faintness of the optical 
counterparts, the lowest identification fractions are achieved in the recent 
deepest samples: 87\% for the {\em XMM--Newton} survey in the Lockman Hole and 88\% 
in the CDF--N. A surprisingly high identification fraction of 98\% has 
been achieved in the CDF--S through the utilization of photometric 
redshifts based on extremely faint optical/NIR photometry.  

For the computation of the soft X-ray luminosity function SXLF, it is important to define the
available survey solid angle as a function of limiting flux. In case there 
is incompleteness in the spectroscopic identifications in the {\em ROSAT}
surveys, we have made the usual assumption that the redshift/classification
distribution of these unidentified sources is the same as the identified 
sources at similar fluxes by defining the 'effective' survey solid angle as the
geometrical survey solid angle multiplied by the completeness of the 
identifications (see Paper I). This assumption is 
not correct when the source is unidentified due to non--random causes, 
in particular its optical faintness. The treatment of this identification
incompleteness and the effect of the optical limit on the derived 
space densities is discussed in detail in Sect~\ref{sec:mxs}.

In several surveys we had to choose an X--ray flux limit a posteriori,
based on optical completeness criteria, i.e. maximising the number of optically
identified sources, while simultaneously minimising the number of unidentified
objects. This procedure can introduce a bias against optically faint sources,
if the reason for the missing redshift is the optical faintness of the source
and in fields with relatively few objects. We have tried to minimise the 
impact of this ''gerrymandering'' effect, e.g. by allowing a number of
unidentified sources to enter the sample and then defining the corresponding
X--ray flux limits in the geometric mean between the last identified and
the next unidentified source. In addition, the wide range for X--ray and
thus optical flux limits in our survey tends to reduce biases, which 
occur at the flux limit of individual surveys. The problem of missing 
redshifts in the faintest surveys is addressed specifically in 
Sect~\ref{sec:exlf} and ~\ref{sec:mxs}.  

\begin{figure*}
   \centering
   \includegraphics[width=12cm]{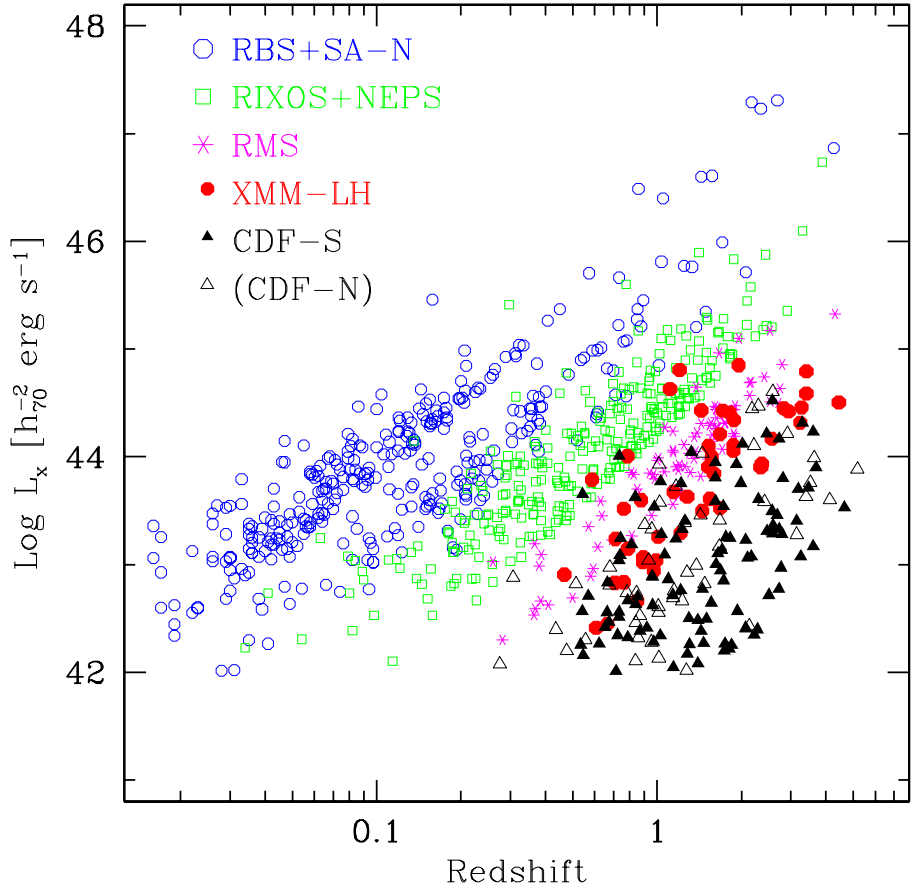}
   \caption{The AGN--1 soft X--ray sample in the $z$-${\rm log}\;L_{\rm x}$
   plane.}
   \label{fig:hubble} 
   \end{figure*}

\begin{figure*}
   \centering
   \includegraphics[width=10cm]{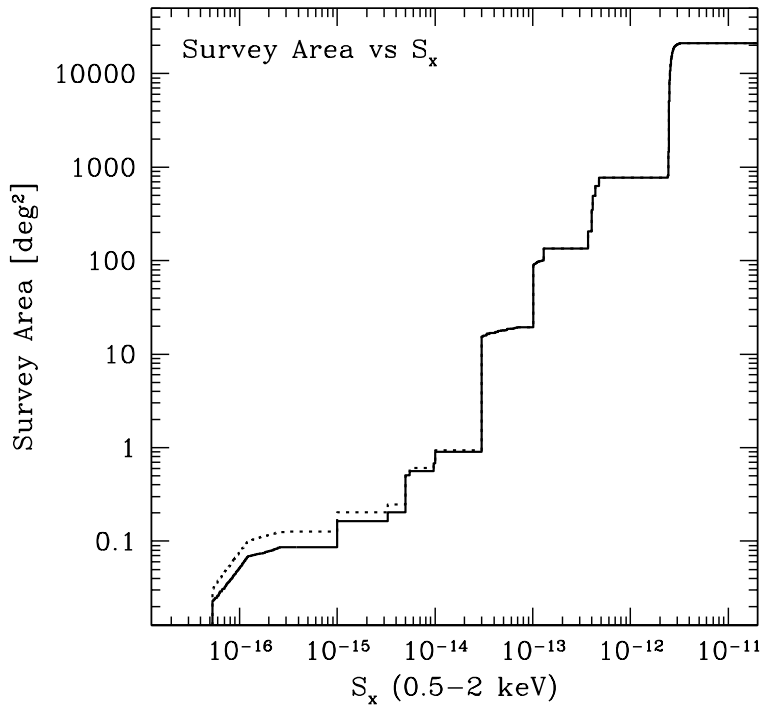}
   \caption{ The survey solid angle of the combined soft X--ray sample
   as a function of flux. The solid line shows the case where 
   the CDF--N sample is excluded as used in Sect.~\ref{sec:exlf}.
   The dotted line includes CDF--N and is used in Sect.~\ref{sec:mxs}.}
   \label{fig:area}
   \end{figure*}

Below we summarize our sample selection and completeness for each survey.  
Figure~\ref{fig:hubble} shows the AGN--1 sample in the redshift -- luminosity plane. 
Figure~\ref{fig:area} gives the combined solid angle versus flux 
curve. Both the sample and the solid angle coverage are available in computer readable form under {\tt http://mpe.mpg.de/\~\ ghasinger}.

%
%
%

\subsection{The {\em ROSAT} Bright Survey (RBS)}

The RBS identified the brightest $\sim$ 2000 X--ray sources
detected in the {\em ROSAT} All--Sky Survey (RASS, Voges et al.
\cite{vog99}) at high galactic latitude, $|b| > 30\degr$, excluding
the Magellanic Clouds and the Virgo cluster, 
with {\em ROSAT} PSPC count rates above 0.2\,s$^{-1}$. This program achieved 
a spectroscopic completeness of 99.5\% (Schwope et al. 
\cite{schw00}). We selected the sub--sample of 931 
sources with count rates above 0.2 s$^{-1}$ in the {\em ROSAT} 
0.5--2 keV band (PSPC channels 52-201), which is 100\% identified. 
Since the absorption in our galaxy varies from place to place,
the same count rate limit corresponds to different 0.5--2 keV flux 
limits based on the different galactic $N_{\rm H}$ values. The 
$N_{\rm H}$ value ranges from $(0.5-16)\times 10^{20}$ 
${\rm cm}^{-2}$ in the RBS survey area. Correspondingly,
the survey solid angle varies steeply with flux from about 3000 deg$^2$ 
at a flux limit of $S_{\rm X14}=246$ to a total of 20391 deg$^2$ 
at a flux limit of $S_{\rm X14}=360$.

\subsection{The RASS Selected-Area Survey North (SA--N)}

This survey gives optical identifications of a representative 
sample of northern ($\delta>-9\degr$) RASS sources in six study areas 
outside the Galactic plane ($|b| > 19.6\degr $) with a total of 
$685 \deg^2$. A count rate limited complete RASS subsample 
comprising 674 sources has been identified (Appenzeller et al.
\cite{app98}). The fields selected for the survey have 
a Galactic column density in the range $N_{\rm H}=(2-11)\times 10^{20}$
${\rm cm}^{-2}$. We have further selected our sample such that 
each of the six fields has a complete {\em ROSAT} hard-band (0.5--2 keV; 
channels 52-201) countrate-limited sample with complete identifications 
($CR_{\rm 0.5-2 keV} >$ 0.01--0.05 ${\rm cts\;s^{-1}}$). To avoid overlap
with the RBS, those sources common in both samples were 
removed from SA--N, yielding a total of 406 sources with 98.5\%
spectroscopic completeness.

\subsection{The {\em ROSAT} North Ecliptic Pole Survey (NEPS)}

The RASS data in a contiguous area of 80.7 deg$^2$ around the North Ecliptic Pole (Galactic latitude $b>29.8\degr$) have been used to construct a survey consisting of 445 X--ray sources detected above a 4$\sigma$ threshold.
Gioia et al. (\cite{gio03}) and Mullis et al. (\cite{mul04}) have identified 99.6\% of these sources and determined redshifts for the extragalactic objects. Since the exposure in the {\em ROSAT} All-Sky Survey increases significantly towards the North-Ecliptic Pole, the actual survey sensitivity is a strong function of ecliptic latitude. The original NEPS sample is selected in the full {\em ROSAT} PSPC band. For consistency with the other surveys used in our work, we selected sources detected significantly in the PSPC hard band (0.5--2 keV; channels 52-201) by specifying a hard count rate limit as a function of ecliptic latitude. New 0.5--2 keV fluxes were calculated from the hard PSPC count rates in the same way as for the RBS and SA--N objects, taking into account the Galactic neutral hydrogen column density varying in the range $2.6-6.2\times10^{20}~cm^{-2}$ across the NEPS region. Due to the large gradient in exposure times, we have cut the sample to 252 sources with fluxes above $S_{\rm X14}= 10.1$, where the solid angle of this survey is 70.5 deg$^2$, increasing to 80.8 deg$^2$ at $S_{\rm X14}= 12.4$. Only one of these sources remains unidentified in the NEPS sample.

\subsection{The {\em ROSAT} International X--ray/Optical Survey (RIXOS)}

The {\em ROSAT} International X--ray/Optical Survey (RIXOS, Mason et al.
\cite{mas00}) is a medium-sensitivity survey and optical identification 
program of X--ray sources discovered in {\em ROSAT} high Galactic latitude fields 
($|b|>28\deg$) and observed with the Position Sensitive Proportional Counter 
(PSPC) detector
with a minimum exposure time of 8 ks. The survey comprises 82 {\em ROSAT} PSPC fields and made use of the central 17 arcmin of each  field, however, excluding the target region for pointings on known X--ray sources. The total survey contains 395 X--ray sources, selected in the PSPC 0.5--2 keV band. A flux limit of $S_{\rm X14}= 3.0$ was adopted for the survey, substantially above the detection threshold of each field, however, the actual spectroscopic completeness
limit varies from field to field. We have chosen a strategy, which on one hand
maximises the sample of identified AGN--1 and on the other hand minimises the number of unidentified sources. There are 51 fields (12.3 deg$^2$) identified completely down to the survey flux limit. Three fields have such a low identification fraction, that we ignore them. For the remaining 28 fields we allow at most one unidentified source. If an unidentified source has the lowest flux of the subsample of a particular field, we exclude this source and raise the flux limit for this field to the geometric average between the flux of this source and that of the last identified source. This way we can define a clean RIXOS sample comprising 340 objects and only 14 unidentified sources, i.e. an identification fraction of 95.9\%. The survey solid angle, corrected for spectroscopic incompleteness, rises from 15.0 deg$^2$ at 
$S_{\rm X14}= 3.0$ to 19.5 deg$^2$ at $S_{\rm X14}= 10.2$.        

\subsection{The {\em ROSAT} Medium Survey (RMS)}

For this work we have grouped a number of medium--deep {\em ROSAT} surveys with flux limits in the range $S_{\rm X14}= 0.5-1$ into the RMS. In particular these comprise pointed observations at the North Ecliptic Pole (Bower et al. \cite{bow96}), the UK Deep Survey (McHardy et al. \cite{mch98}), the Marano field (Zamorani et al. \cite{zam99}) and the outer parts of the Lockman Hole (Schmidt et al. \cite{schm98}, Lehmann et al. \cite{leh00}). The North Ecliptic pole pointing covers the same sky area as the center of the NEPS, however, to a flux limit of $S_{\rm X14}= 1.0$. Again, we remove the overlapping sources between the two surveys. For the UK Deep Survey and the Marano Field we define a flux limit of $S_{\rm X14}= 0.5$, following Paper I. For the {\em ROSAT} PSPC survey of the Lockman Hole we only chose the region not covered by the deeper RDS/XMM survey (see section~\ref{sec:xmmlh}) but otherwise selected the completely identified sample with the same flux limits as those chosen for the {\it ROSAT Ultradeep Survey} UDS by Lehmann et al. \cite{leh01}: $S_{\rm X14}> 0.96$ for PSPC off--axis angles in the range 12.5--18.5 arcmin and $S_{\rm X14}> 0.55$ for off--axis angles smaller than 12.5 arcmin. Overall, the RMS contains 124 sources, at an identification completeness of 94.4\%. Correspondingly, the corrected survey solid angle varies in the range 0.30--0.70 deg$^2$ for flux limits $S_{\rm X14}$=0.5--1.
  
\subsection{Deep {\em XMM--Newton} survey of the Lockman Hole (XMM/RDS)}
\label{sec:xmmlh}

The Lockmam Hole (XMM/RDS) has been observed by {\em XMM--Newton} a total of 17 times during the PV, AO--1 and AO--2 phases of the mission, with total good exposure times in the range 680--880 ks in the PN and MOS instruments (see Hasinger et al. \cite{has01,has04} and Worsley et al. \cite {wor04} for details). Spectroscopic
optical identifications of the {\em ROSAT} sources in the LH have been presented 
by Schmidt et al. (\cite{schm98}) and Lehmann et al. (\cite{leh00,leh01}) and 
a new catalogue from the {\em XMM--Newton} PV phase is given in Mainieri et al. (\cite{mai02}). Some photometric redshifts have been discussed in Fadda et al., (\cite{fad02}). Here we selected sources from the 770 ksec dataset (Brunner et al., 2005, in prep.) with additional spectroscopic identifications obtained 
with the DEIMOS spectrograph on the Keck telescope in spring 2003 and 2004
by M. Schmidt and P. Henry (Szokoly al., 2005, in prep.). In order to 
maximise the spectroscopic/photometric completeness of the sample, we selected objects in two off--axis intervals: $S_{\rm X14}= 0.38$ for off--axis angles in the range 10.0--12.5 arcmin and $S_{\rm X14}= 0.13$ for off--axis angles smaller than 10.0 arcmin. The total number of sources in the XMM/RDS is 81, with 8 potential AGN--1 still unidentified.

\subsection{{\em Chandra} Deep Field South (CDF--S)}\label{sec:cdfs}

We have used the catalogue of Giacconi et al. (\cite{gia02}) based 
on the 1 Ms observation of the CDF--S (Rosati et al. \cite{ros02}). Spectroscopic identifications with the FORS instruments at the ESO VLT have been obtained by Szokoly et al. (\cite{szo04}), yielding a spectroscopic completeness around 60\%.
Additional spectroscopic redshifts of CDF--S X--ray sources have been obtained with the VIMOS spectrograph at the ESO VLT (Lefevre et al., 2004). The field is also
included in the COMBO-17 intermediate--band optical survey, which gives very reliable photometric redshifts for the brighter sources (Wolf et al., \cite{combo17}).  
Very deep NIR photometry has been obtained with the ISAAC camera at the VLT in conjunction with deep optical imaging with the HST ACS as part of the GOODS project (Dickinson \& Giavalisco \cite{dic03}, Mobasher et al. \cite{mob04}).  The CDF--S therefore offers the highest quality photometric redshifts of faint 
X--ray sources, which are discussed in Zheng et al. (\cite{zhe04}) and Mainieri et al. (\cite{mai05}). Using the ISAAC images, tentative photometric redshifts  could even be assigned to several of the extreme X--ray/optical sources (EXOs) discussed by Koekemoer et al. (\cite{koe04}). We selected all sources from 
the Giacconi et al. \cite{gia02} catalogue within 10 arcmin from the {\em Chandra} pointing center significantly detected in the 0.5--2 keV band. The  sample thus contains a total of 293 objects. Combining all spectroscopic and photometric redshifts,
only 2 sources in the CDFS  remain unidentified, of which one could be an AGN--1.  The survey solid angle for the CDF--S has been estimated using a simple off--axis dependent flux limit. The solid angle increases from 0.023 deg$^2$ at $S_{\rm X14}= 0.0053$ to 0.087 deg$^2$ at $S_{\rm X14}= 0.027$.

\subsection{{\em Chandra} Deep Field North (CDF--N)}
We have used selected X--ray sources from the 2 Ms CDF--N
source catalogue by Alexander et al. (\cite{ale03}) along with 
optical  identifications by Barger et al. (\cite{bar03}) for our
AGN--1 sample. Following Szokoly et al. (\cite{szo04}), 
we selected AGN--1 either from broad permitted Balmer lines 
or from the X--ray luminosity and hardness, using HR$<$--0.2. 
We set our flux limits such that we also have sufficient hard (2--8 keV) 
sensitivity to exclude objects with HR$>$--0.2 and to include
as many sources as possible which meet these criteria.
 Unlike the CDF--S, the CDF--N exposure map has a complicated 
structure and a simple off--axis dependence of the limiting flux is 
not a good approximation.  Thus we have used the rectangular region
of 170 arcmin$^2$, which has the deepest coverage and is mostly co--spatial
with the region covered by HST ACS in the GOODS project 
(Giavalisco et al. \cite{goods1}, Cowie et al. \cite{cowie04}). 
Within this region, the exposure and background are smooth enough that 
the photon counts limit of the detected sources can be approximated by a simple
function of off--axis angle. In practice, due to statistical fluctuations, three sources have 
{\em upper limits} to HR between --0.1 and --0.2 and those have been considered
to meet our hardness ratio criterion. Among the 128 sources meeting 
the soft counts limit and hardness ratio criteria, 20 are unidentified 
and 5 are stars (85\% completeness). Only one broad--line AGN had a 
harder hardness ratio than our limit; this was also included
in our type--1 AGN sample. 
The flux--solid angle relation has been calculated from the ``limiting 
flux map'', where the counts limit is divided
by the soft--band exposure map (in seconds) and multiplied by
the conversion factor of $5\cdot10^{-12}\;{\rm erg\,s^{-1}\,cm^{-2}}$
(Alexander et al., \cite{ale03}). Due to the incompleteness in this field, 
where most
unidentified sources are optically faint, this sample has not been 
included in the analysis in Section~\ref{sec:exlf},
but considered in Section~\ref{sec:mxs}, where a method is developed to
account for the optical magnitude limit in calculating the survey
volume.   

%

%

   \begin{figure*}
   \centering
   \includegraphics[width=12cm]{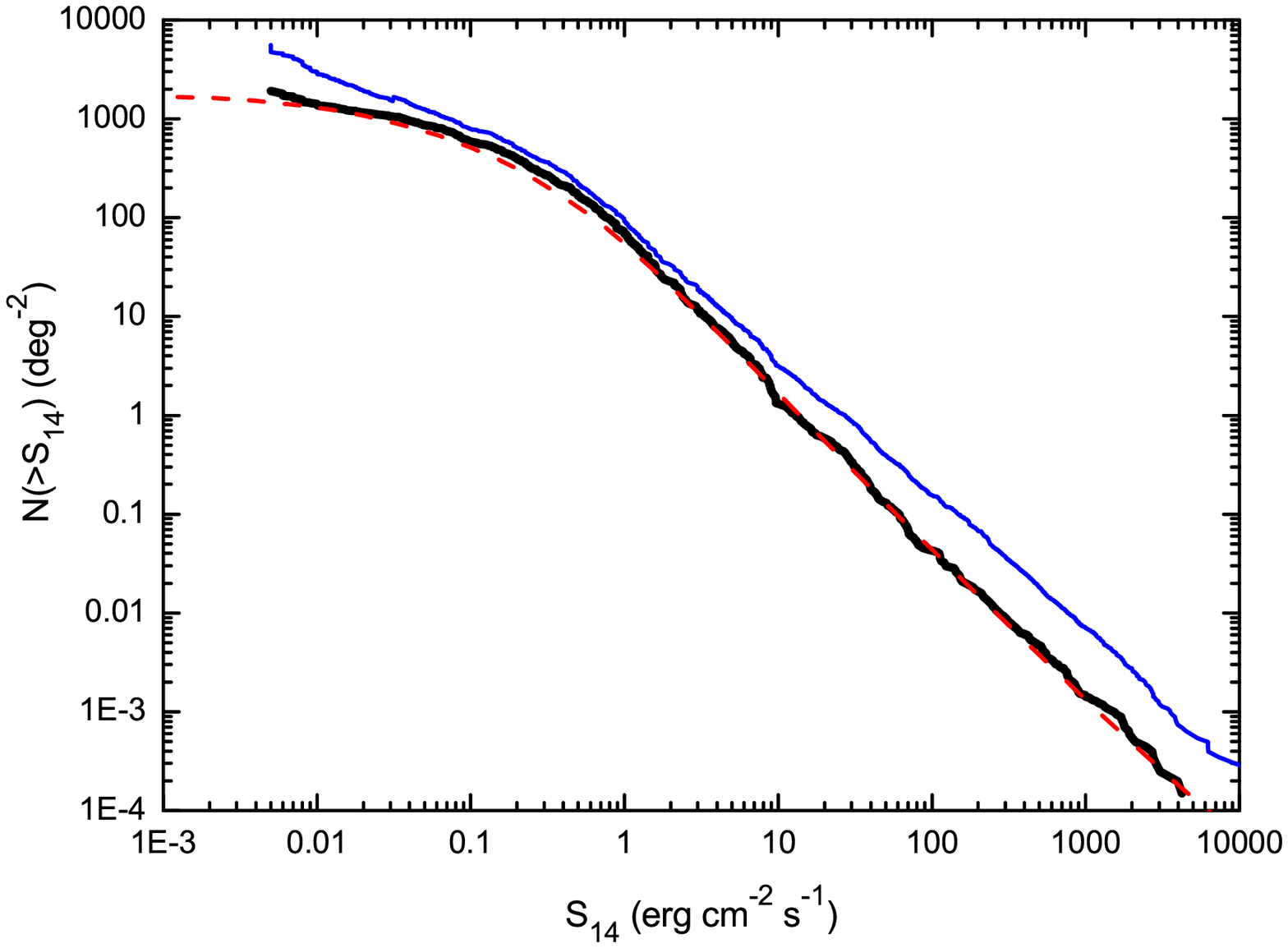}
   \includegraphics[width=12cm]{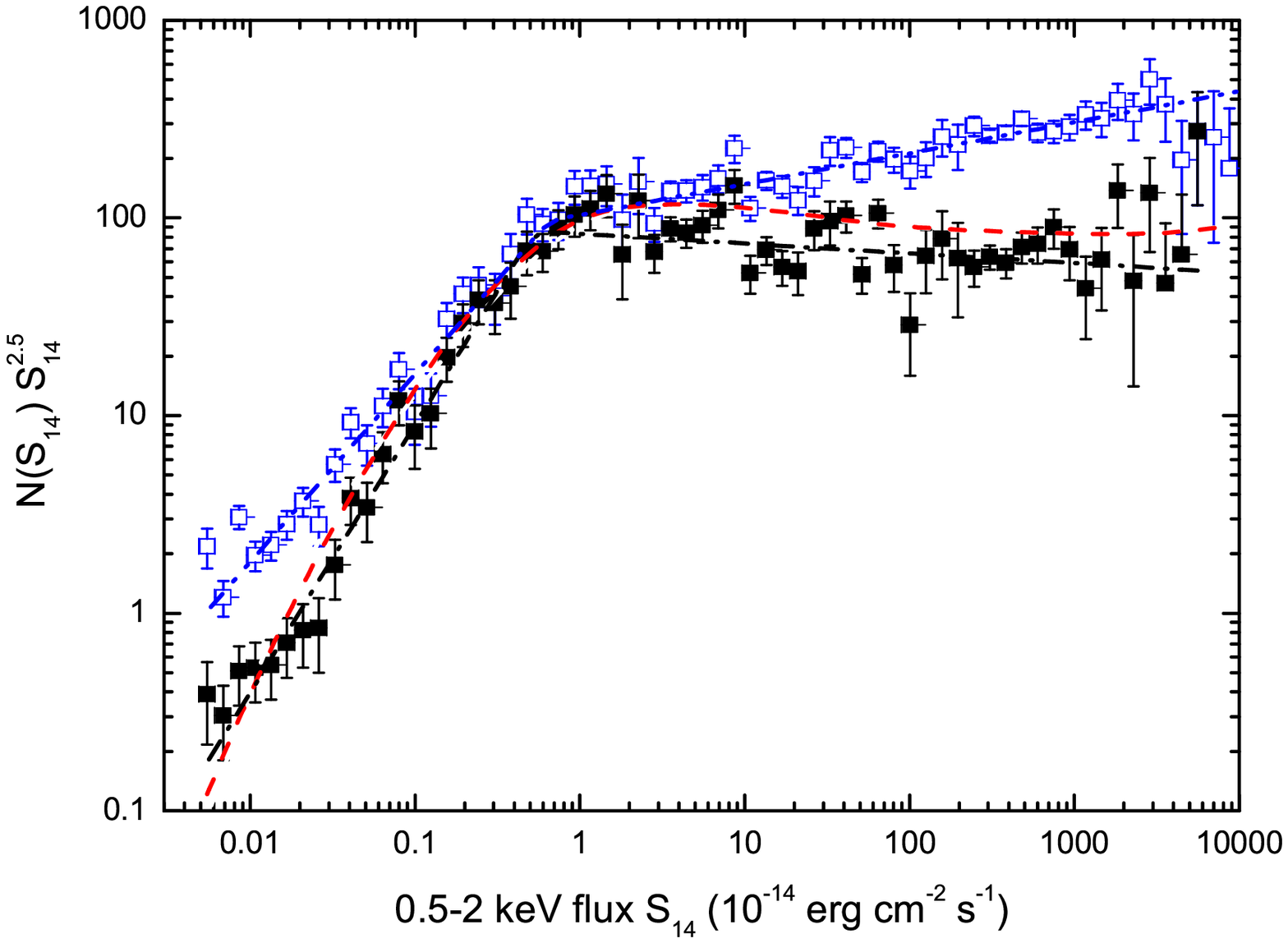}
   \caption{(a) Cumulative number counts N($>$S) for the total sample
   (upper thin line) and the AGN--1 subsample (lower thick line).
(b) Differential number counts of the total sample of X--ray sources
   (open squares) and the AGN--1 subsample (filled squares).  
   The dot-dashed lines refer to broken powerlaw fits to the differential
   source counts (see text). The dashed red line shows 
   the prediction for type--1 AGN based on the model described in 
   in section~\ref{sec:mxs}. }
   \label{fig:ns} 
   \end{figure*}
%
%

\section{Number counts for different source classes}
\label{sec:lns}

The combination of a large number of surveys with a wide range of sensitivity
limits and solid angle coverage presents a unique resource. On one hand, the surveys presented here resolve the soft X--ray background almost completely. On the other hand, we have an almost complete optical identification and redshift determination for all constituents. For the first time in any astrophysical waveband we are thus in the position to study the complete contribution of different object classes to the X--ray background and their evolution with cosmic time.
Using the solid angle versus flux limit curve given in Figure~\ref{fig:area} we compiled number counts for the total sample including all classes of sources and for the subclass of AGN--1. Figure~\ref{fig:ns} (top) shows the cumulative source counts. For clarity we also show normalized differential source counts $dN/dS_{\rm X14} S_{\rm X14}^{2.5}$ in the bottom panel of Figure~\ref{fig:ns}. Euclidean source counts would correspond to horizontal lines in this graph. 

For the total source counts, the well-known broken power law behaviour is confirmed with high precision. We fitted a broken power law to the differential source counts and obtain power law indices of $\alpha_b=2.34\pm0.01$ and $\alpha_f=1.55\pm0.04$ for the bright and faint end, respectively, a break flux of $S_{\rm X14}= 0.65\pm0.10$ and a normalisation of  dN/d$S_{\rm X14} = 103.5 \pm 5.3$ deg$^{-2}$ at $S_{\rm X14} = 1.0$ with a reduced $\chi^2$=1.51. The total differential source counts, normalized to a Euclidean behaviour (dN/d$S_{\rm X14} \times S_{\rm X14}^{2.5}$ is shown with open symbols in Figure~\ref{fig:ns}. We see that the total source counts at bright fluxes, as determined by the {\em ROSAT} All-Sky Survey data, are significantly flatter than Euclidean, consistent with the discussion in Hasinger et al. (\cite{has93}). Moretti et al. (\cite{mor03}), on the other hand, have derived a significantly steeper bright flux slope ($\alpha_b \approx 2.8$) from {\em ROSAT} HRI pointed observations. This discrepancy can probably be attributed to the selection bias against bright sources, when using pointed observations where the target area has to be excised.

Type--1 AGN are the most abundant population of soft X--ray sources. For the determination of the AGN--1 number counts we include those unidentified sources, which have hardness ratios consistent with AGN--1 (a contribution of $\sim 6\%$, see Table~\ref{tab:samp}). Figure~\ref{fig:ns} shows, that the break in the total source counts at intermediate fluxes is produced by type--1 AGN, which are the dominant population there. Both at bright fluxes and at the faintest fluxes, type--1 AGN contribute about 30\% of the X--ray source population. At bright fluxes, they have to share with clusters, stars and BL-Lac objects, at faint fluxes they compete with type--2 AGN and normal galaxies. We fitted a broken power law to the differential AGN--1 source counts and obtain power law indices of $\alpha_b=2.55\pm0.02$ and $\alpha_f=1.15\pm0.05$ for the bright and faint end, respectively, a break flux of $S_{\rm X14}= 0.53\pm0.05$, consistent with that of the total source counts within errors, and a normalisation of of dN/d$S_{\rm X14} = 83.2 \pm 5.5$ deg$^{-2}$ at $S_{\rm X14} = 1.0$ with a reduced $\chi^2$=1.26. The AGN--1 differential source counts, normalized to a Euclidean behaviour (dN/d$S_{\rm X14} \times S_{\rm X14}^{2.5}$) is shown with filled symbols in Figure~\ref{fig:ns}. Also shown are the predictions of the best--fit SXLF models discussed in Section~\ref{sec:mxs}.

\section{The SXLF and the Space Density Function}
\label{sec:exlf}

\subsection{Basic method}
\label{sec:exlf1}

 In this section, we present the binned {\it Soft X--ray Luminosity Function}
(SXLF) of type--1 AGNs. The basic approach is to use the $N^{\rm obs}/N^{\rm mdl}$ estimator 
described in Paper II. The procedure is outlined below:
\begin{enumerate}
\item Divide the combined sample into several redshift shells. 
  For each redshift shell, fit the AGN XLF with a smooth analytical
  function using a Maximum-likelihood fit over each object (i.e.,
  without binning; see Paper I for details). 
\item For the fitted model in each redshift shell, check the absolute 
  goodness of fit with one-- and two--dimensional Kolmogorov--Smirnov tests 
  (hereafter, 1D--KS and 2D--KS tests respectively; 
  Press et al. \cite{numrec}, Fassano \& Franceschini \cite{ff_2dks}). 
  The K--S tests are also for the unbinned data sets and 
  thus are free from artefacts and biases due to binning.
\item For each redshift shell, bin the objects in luminosity 
  bins to determine the observed number of objects ($N^{\rm obs}$).
\item For each luminosity bin, evaluate the analytical
  fit at the central luminosity/redshift 
 ($d\Phi^{\rm mdl}/d{\rm log} L_{\rm x}$).
\item Calculate the predicted number of AGNs in the bin ($N^{\rm mdl}$).
\item The final result is 

  \begin{equation}
  d\Phi/d\,{\rm log\,}L_{\rm x}=d\Phi^{\rm mdl}/d\;{\rm log\,}L_{\rm x} \cdot
  \;N^{\rm obs}/N^{\rm mdl} 
  \end{equation}

\end{enumerate}

 For the analytical expression of the SXLF in each redshift shell, we use 
the {\em smoothed two power law} formula. Because the redshift shells
have a finite widths, the fit results depend on the evolution of the 
SXLF within them: 
\begin{equation}
\frac{{\rm d}\;\Phi\,(L_{\rm x},z) }{{\rm d\;log}\;L_{\rm x}}
 \propto \left[\left(\frac{L_{\rm x}}{{L_{\rm x,*}}}\right)^{{\gamma_1}}
         +\left(\frac{L_{\rm x}}{{L_{\rm x,*}}}\right)^{{\gamma_2}}
  	\right]^{-1} \cdot e_{\rm d}(z,L_{\rm x}),
\label{eq:2po}
\end{equation}
where $e_{\rm d}(z,L_{\rm x})$ is the density evolution factor.
While the final results are insensitive to the detailed behavior of  
$e_{\rm d}(z,L_{\rm x})$ within the shell at most locations in the 
$(L_{\rm x},z)$ space, we have taken our best-estimate by using the
luminosity-dependent density evolution (LDDE) model derived later in 
Sect. \ref{sec:pleldde}. The luminosity range of the fit is from 
log$\,L_{\rm x}$=42.0 to the maximum available luminosity in the sample.

 In this section, we tried to make the sample as complete as possible,
and we excluded the CDF--N from the analysis, where the incompleteness
fraction is significant and most of the unidentified sources are
optically faint. All of the unidentified sources in the {\em ROSAT} samples
are optically bright and the reasons for them to be unidentified
are mostly by random causes, i.e., are  not correlated with the 
intrinsic properties of the source. 
 For the CDF--S, extensive photometric redshift studies including 
COMBO-17 (Wolf et al., \cite{combo17}) and a careful individual 
photometric redshift determination of X--ray sources by Zheng et al. \cite{zhe04} 
and Mainieri et al., \cite{mai05} has left only one potential AGN--1 without 
redshift information. For RDS/XMM, 2 of the 8 unidentified 
sources which could be type--1 AGNs from X--ray hardness/spectra criteria
are optically bright and they have remained unidentified so far for random
reasons. The remaining 6 are optically faint ($R\geq 24.0$) and the reason
for remaining unidentified may well be correlated with redshift.
 
As our nominal case, we took the first--order approach 
and defined ``effective'' survey solid angle (as a function of flux), 
which is the geometrical survey solid angle multiplied by the completeness, 
i.e. the fraction of identified X--ray sources in the survey, whether
or not they are optically faint or bright. The correction has been made in 
each survey.  In addition,  we have also considered the upper bounds 
on the binned SXLF and the space density function from the sample where all the 
unidentified optically faint ($R\geq 24.0$) sources in turn are assigned 
the central redshift of each redshift shell.

In the latter case, we used the geometrical solid angle for the CDF--S 
and the incompleteness correction to the RDS/XMM solid angle was
only for the optically bright $R<24$ unidentified sources.  
    
\subsection{The binned SXLF}\label{sec:xlf}

The best fit parameters of Eq.~\ref{eq:2po} for each redshift shell are 
shown in Table~\ref{tab:2po} along with the results of the 1D-- and 2D--KS 
tests (see the notes of the table). The normalization is defined by:
\begin{equation}
A_{\rm 44}=\frac{{\rm d}\;\Phi\,(L_{\rm x}=10^{44}\;{\rm erg\;s^{-1}} ,z=z_{\rm c}) }
{{\rm d\;log}\;L_{\rm x}},
\end{equation}   
where $z_{\rm c}$ is the central redshift of the shell.
The parameter errors
in Table~\ref{tab:2po} correspond to a likelihood change of 2.7 
(90\% confidence errors), except for the normalization $A_{44}$, which cannot 
be a fit parameter in the maximum-likelihood method. The errors 
of $A_{44}$ are simply taken as the 90\% Poisson errors of the number of 
the sources. Defining the normalization at a fixed luminosity 
($\log L_{\rm x}=44$) minimizes its dependence on other parameters. 
In any case, Eq.~\ref{eq:2po} gives a statistically satisfactory 
expression for all redshift shells as shown in Table~\ref{tab:2po}.  

\begin{table*}
\caption[]{Best--fit parameters for each redshift shell}
\begin{center}
\begin{tabular}{ccrccccc}
\hline\hline
$z$-range & $z_{\rm c}$ & N & $A_{44}^{\rm a}$ & $\log\; L_{\rm x,*}^{\rm a}$
  & $\gamma_1$ & $\gamma_2$ & KS--prob$^{\rm b}$ \\
\hline                                               
0.0-0.2 &  0.1 & 268 &$(3.64\pm0.22)\cdot 10^{-7}$ &
43.45$^{+0.32}_{-0.27}$ &  0.35$^{+0.31}_{-0.40}$ &
  2.1$^{+0.4}_{-0.3}$ & 0.84,0.17,0.14\\
0.2-0.4 &  0.3 & 139 &$(9.76\pm0.83)\cdot 10^{-7}$ &
43.91$^{+0.38}_{-0.36}$ &  0.70$^{+0.31}_{-0.45}$ &
  2.5$^{+0.5}_{-0.3}$ & 0.97,0.71,0.65\\
0.4-0.8 &  0.6 & 143 &$(1.84\pm0.15)\cdot 10^{-6}$ &
43.91$^{+0.41}_{-0.41}$ &  0.84$^{+0.23}_{-0.32}$ &
  2.3$^{+0.3}_{-0.2}$ & 1.00,0.03,0.08\\
0.8-1.6 &  1.2 & 187 &$(1.19\pm0.09)\cdot 10^{-5}$ &
43.97$^{+0.13}_{-0.13}$ &  0.10$^{+0.17}_{-0.19}$ &
  2.3$^{+0.2}_{-0.1}$ & 0.91,0.68,0.88\\
1.6-3.2 &  2.4 & 110 &$(1.03\pm0.10)\cdot 10^{-5}$ &
44.39$^{+0.16}_{-0.17}$ &  0.15$^{+0.16}_{-0.18}$ &
  2.3$^{+0.2}_{-0.2}$ & 0.91,0.55,0.44\\
3.2-4.8 &  4.0 &  17 &$(4.53\pm1.10)\cdot 10^{-6}$ &
44.43$^{+0.40}_{-0.37}$ & -0.26$^{+0.56}_{-0.75}$ &
  1.9$^{+0.4}_{-0.3}$ & 0.95,0.21,0.39\\
\hline
\end{tabular}
\end{center}
\label{tab:2po}
Parameter values which have been fixed during the fit are labelled
by '(*)'.  $^{\rm a}$Units -- $A_{44}$: $h_{70}^3\;{\rm Mpc^{-3}}$,\,\,
$L_{\rm x,*}$: $10^{44}\;h_{70}^{-2}{\rm erg\;s^{-1}}$.
$^{\rm b}$ The three values are probabilities in two
1D--KS test for the distribution, $L_{\rm x}$, 1D--KS test for the
$z$ distribution and the 2D--KS test for the ($L_{\rm x}$,$z$) space respectively.
\end{table*}

\begin{figure*}
\centering
\includegraphics[width=18cm]{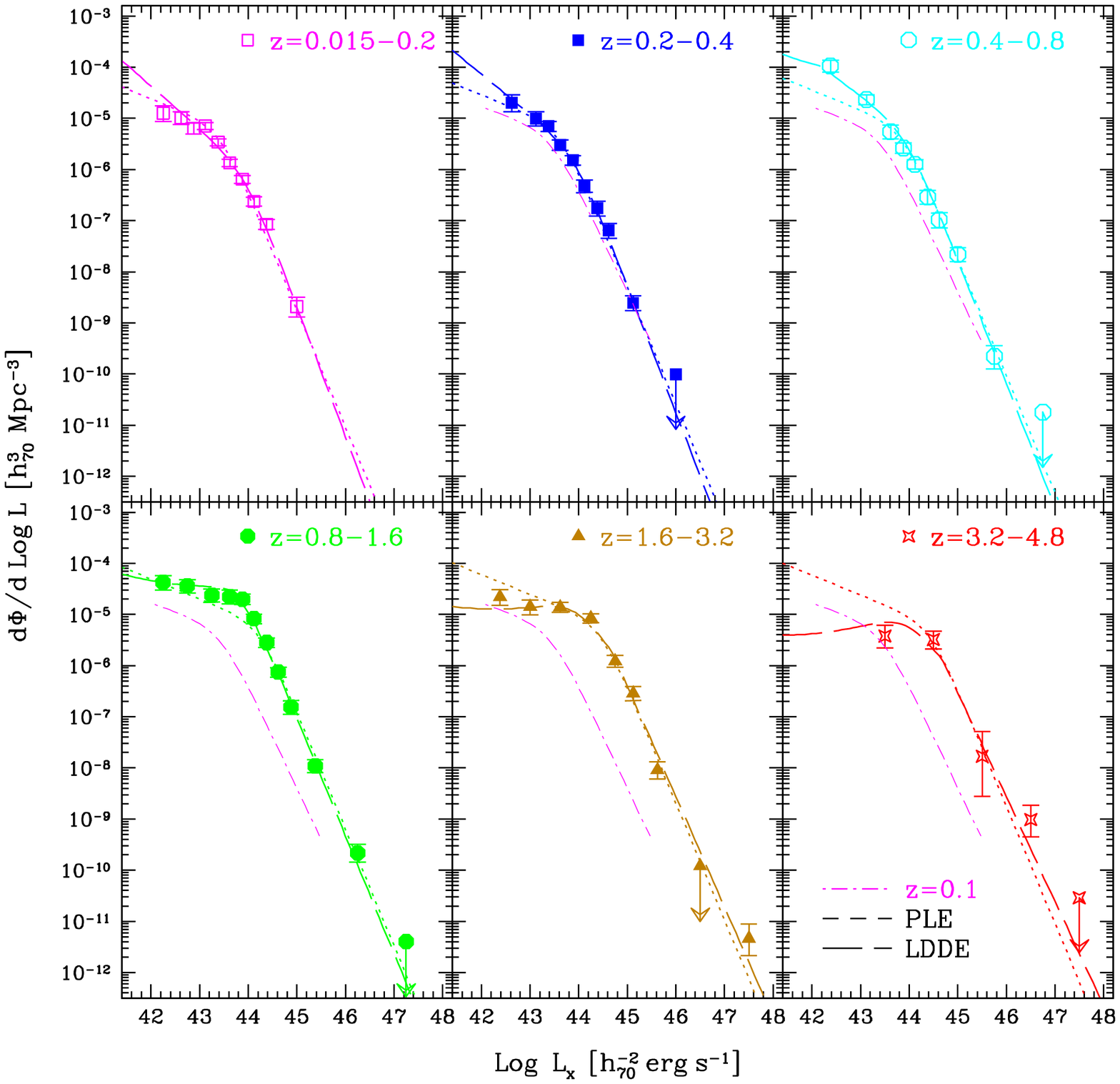}
\caption{The soft X--ray luminosity function
 of the type--1 AGN sample in different redshift shells for the
 nominal case as labeled. The error bars correspond to 68\% Poisson 
 errors of the number of AGNs in the bin. The best--fit two power law
 model for the $0.015<z<0.2$ shell are overplotted in the higher
 redshift panels for reference. The dotted and long-dashed 
 lines give the best--fit PLE and LDDE models discussed in 
 Sect.~\ref{sec:pleldde} respectively.}
\label{fig:exlf}
\end{figure*}

We have made luminosity bins starting with a minimum luminosity of 
$\log\; L_{\rm x}=42.0$ with a smallest bin size of 
$\Delta \log\; L_{\rm x}=0.25$ in each redshift shell. If there 
are fewer than 10 AGNs in a bin, we have further rebinned up to a 
maximum bin size of $\Delta {\rm log\;} L_{\rm x}=1.0$. 
Table~\ref{tab:exlf} shows the full binned results for the nominal case,
along with observed number of AGNs ($N^{\rm obs}$), model
(Table~\ref{tab:2po}) predicted number ($N^{\rm mdl}$) and final estimated
values SXLF value at the center of each bin. For reference, the additional
number $N^f$ of AGNs for the case where all the optically-faint unidentified
sources are assigned the central redshift of the bin (in duplicate,
as described above) are also shown in the last column of Table~\ref{tab:exlf}.
The full SXLF in the 6 redshift shells is plotted in 
Figs.~\ref{fig:exlf} in separate panels. In all but the closest redshift
shell panel, the best--fit two power law function to the $0.015<z<0.2$ 
SXLF (Table~\ref{tab:2po}) are also overplotted for reference. Three 
overall analytical expressions discussed in Sect.~\ref{sec:pleldde} 
are also overplotted for comparison as discussed there.
 Because of the high completeness of our sample, the redshift distribution of the 
optically faint sources affects the final SXLF results very little except for the 
case where all of them happen to fall into
the highest redshift shell. In this case, the SXLF in the  $3.2<z<4.8$,
$44<{\rm log}\;L_{\rm x}$ bin almost double.  This is also verified by a
comparison with the alternative approach outlined in 
Sect.~\ref{sec:mxs}.

\begin{table*}
\caption[]{The full binned SXLF values}
\begin{center}
\begin{tabular}{ccrrcc}
\hline\hline
$z$ & log\,{$L_{\rm x}$}$^{\rm a}$
  & $N^{\rm obs}$ & $N^{\rm mdl}$ &
     $\frac{{\rm d \Phi}}{{\rm d\, log}\,L_{\rm x}}^{\rm b}$ &
     $(N^{\rm f})$\\

\hline
0.015 - 0.2 & 42.00 - 42.50 &  10 &  11.5  
 &  $(1.3^{+0.5}_{-0.4})\cdot 10^{-5}$ \\ 
0.015 - 0.2 & 42.50 - 42.75 &  15 &  15.3 
 &  $(1.0^{+0.3}_{-0.3})\cdot 10^{-5}$ \\ 
0.015 - 0.2 & 42.75 - 43.00 &  20 &  25.2 
 &  $(6.4^{+1.7}_{-1.4})\cdot 10^{-6}$ \\ 
0.015 - 0.2 & 43.00 - 43.25 &  48 &  38.6 
 &  $(7.1^{+1.1}_{-1.0})\cdot 10^{-6}$ \\ 
0.015 - 0.2 & 43.25 - 43.50 &  45 &  44.9 
 &  $(3.4^{+0.6}_{-0.5})\cdot 10^{-6}$ \\ 
0.015 - 0.2 & 43.50 - 43.75 &  38 &  45.7 
 &  $(1.4^{+0.2}_{-0.2})\cdot 10^{-6}$ \\ 
0.015 - 0.2 & 43.75 - 44.00 &  36 &  34.6 
 &  $(6.5^{+1.2}_{-1.1})\cdot 10^{-7}$ \\ 
0.015 - 0.2 & 44.00 - 44.25 &  27 &  23.7 
 &  $(2.4^{+0.5}_{-0.5})\cdot 10^{-7}$ \\ 
0.015 - 0.2 & 44.25 - 44.50 &  22 &  16.9 
 &  $(8.5^{+2.1}_{-1.8})\cdot 10^{-8}$ \\ 
0.015 - 0.2 & 44.50 - 45.50 &   7 &  11.2 
 &  $(2.1^{+1.1}_{-0.8})\cdot 10^{-9}$ \\ 
\\                                        
0.2 - 0.4 & 42.25 - 43.00 &   9 &   9.6   
 &  $(2.0^{+0.9}_{-0.7})\cdot 10^{-5}$ \\ 
0.2 - 0.4 & 43.00 - 43.25 &  12 &  11.2   
 &  $(10.0^{+3.6}_{-2.8})\cdot 10^{-6}$ \\
0.2 - 0.4 & 43.25 - 43.50 &  23 &  19.1   
 &  $(7.0^{+1.7}_{-1.5})\cdot 10^{-6}$ \\ 
0.2 - 0.4 & 43.50 - 43.75 &  22 &  24.1   
 &  $(3.0^{+0.8}_{-0.6})\cdot 10^{-6}$ \\ 
0.2 - 0.4 & 43.75 - 44.00 &  25 &  25.7   
 &  $(1.5^{+0.4}_{-0.3})\cdot 10^{-6}$ \\ 
0.2 - 0.4 & 44.00 - 44.25 &  15 &  18.1   
 &  $(4.7^{+1.5}_{-1.2})\cdot 10^{-7}$ \\ 
0.2 - 0.4 & 44.25 - 44.50 &  11 &  10.6   
 &  $(1.8^{+0.7}_{-0.5})\cdot 10^{-7}$ \\ 
0.2 - 0.4 & 44.50 - 44.75 &  11 &   7.6   
 &  $(6.4^{+2.4}_{-1.9})\cdot 10^{-8}$ \\ 
0.2 - 0.4 & 44.75 - 45.50 &  11 &  11.8    
 &  $(2.5^{+0.9}_{-0.7})\cdot 10^{-9}$ \\  
0.2 - 0.4 & 45.50 - 46.50 &   0 &   0.4    
 &  $< 9.9\cdot 10^{-11}$ \\               
\\                                       
0.4 - 0.8 & 42.00 - 42.75 &  18 &  17.1    
 &  $(1.1^{+0.3}_{-0.2})\cdot 10^{-4}$ &(6)\\ 
0.4 - 0.8 & 42.75 - 43.50 &  23 &  22.0    
 &  $(2.3^{+0.6}_{-0.5})\cdot 10^{-5}$ \\ 
0.4 - 0.8 & 43.50 - 43.75 &  13 &  15.4   
 &  $(5.5^{+1.9}_{-1.5})\cdot 10^{-6}$ \\ 
0.4 - 0.8 & 43.75 - 44.00 &  21 &  23.4   
 &  $(2.6^{+0.7}_{-0.6})\cdot 10^{-6}$ \\ 
0.4 - 0.8 & 44.00 - 44.25 &  26 &  22.7   
 &  $(1.3^{+0.3}_{-0.2})\cdot 10^{-6}$ \\ 
0.4 - 0.8 & 44.25 - 44.50 &  13 &  16.0   
 &  $(2.9^{+1.0}_{-0.8})\cdot 10^{-7}$ \\ 
0.4 - 0.8 & 44.50 - 44.75 &  11 &  11.1   
 &  $(1.0^{+0.4}_{-0.3})\cdot 10^{-7}$ \\ 
0.4 - 0.8 & 44.75 - 45.25 &  13 &   8.8   
 &  $(2.2^{+0.7}_{-0.6})\cdot 10^{-8}$ \\ 
0.4 - 0.8 & 45.25 - 46.25 &   5 &   6.2    
 &  $(2.2^{+1.4}_{-1.0})\cdot 10^{-10}$\\ 
0.4 - 0.8 & 46.25 - 47.25 &   0 &   0.2    
 &  $< 1.8\cdot 10^{-11}$ \\               
\\                                         
0.8 - 1.6 & 42.00 - 42.50 &  12 &  11.0    
 &  $(4.2^{+1.5}_{-1.2})\cdot 10^{-5}$ &(1)\\
0.8 - 1.6 & 42.50 - 43.00 &  13 &  12.4    
 &  $(3.6^{+1.2}_{-1.0})\cdot 10^{-5}$ \\
0.8 - 1.6 & 43.00 - 43.50 &  14 &  17.7    
 &  $(2.4^{+0.8}_{-0.6})\cdot 10^{-5}$ &(6)\\ 
0.8 - 1.6 & 43.50 - 43.75 &  12 &  12.9   
 &  $(2.2^{+0.8}_{-0.6})\cdot 10^{-5}$ \\ 
0.8 - 1.6 & 43.75 - 44.00 &  20 &  16.4   
 &  $(2.0^{+0.5}_{-0.4})\cdot 10^{-5}$ \\ 
0.8 - 1.6 & 44.00 - 44.25 &  27 &  25.5   
 &  $(8.2^{+1.8}_{-1.6})\cdot 10^{-6}$ \\ 
0.8 - 1.6 & 44.25 - 44.50 &  27 &  25.0   
 &  $(2.8^{+0.6}_{-0.5})\cdot 10^{-6}$ \\ 
0.8 - 1.6 & 44.50 - 44.75 &  26 &  25.7   
 &  $(7.5^{+1.7}_{-1.5})\cdot 10^{-7}$ \\ 
0.8 - 1.6 & 44.75 - 45.00 &  13 &  16.7   
 &  $(1.5^{+0.5}_{-0.4})\cdot 10^{-7}$ \\ 
0.8 - 1.6 & 45.00 - 45.75 &  15 &  18.5   
 &  $(1.1^{+0.3}_{-0.3})\cdot 10^{-8}$ \\ 
0.8 - 1.6 & 45.75 - 46.75 &   8 &   4.4    
 &  $(2.2^{+1.0}_{-0.8})\cdot 10^{-10}$\\ 
0.8 - 1.6 & 46.75 - 47.75 &   0 &   0.3    
 &  $< 4.0\cdot 10^{-12}$ \\               
\\                                        
1.6 - 3.2 & 42.00 - 42.75 &  10 &   9.5    
 &  $(2.2^{+0.9}_{-0.7})\cdot 10^{-5}$ \\  
1.6 - 3.2 & 42.75 - 43.25 &  11 &  13.1    
 &  $(1.4^{+0.5}_{-0.4})\cdot 10^{-5}$ & (1)\\  
1.6 - 3.2 & 43.25 - 44.00 &  23 &  22.0    
 &  $(1.4^{+0.3}_{-0.3})\cdot 10^{-5}$ & (6)\\  
1.6 - 3.2 & 44.00 - 44.50 &  27 &  23.0    
 &  $(8.4^{+1.9}_{-1.6})\cdot 10^{-6}$ \\  
1.6 - 3.2 & 44.50 - 45.00 &  16 &  16.9    
 &  $(1.2^{+0.4}_{-0.3})\cdot 10^{-6}$ \\  
1.6 - 3.2 & 45.00 - 45.25 &  12 &   8.4    
 &  $(2.9^{+1.0}_{-0.8})\cdot 10^{-7}$ \\  
1.6 - 3.2 & 45.25 - 46.00 &   8 &  13.1    
 &  $(9.1^{+4.2}_{-3.1})\cdot 10^{-9}$ \\  
1.6 - 3.2 & 46.00 - 47.00 &   0 &   2.9    
 &  $< 1.2\cdot 10^{-10}$ \\               
1.6 - 3.2 & 47.00 - 48.00 &   3 &   0.5    
 &  $(4.6^{+4.2}_{-2.5})\cdot 10^{-12}$ \\ 
\\                                         
3.2 - 4.8 & 43.00 - 44.00 &   5 &   4.9    
 &  $(3.8^{+2.4}_{-1.6})\cdot 10^{-6}$ & (1)\\ 
3.2 - 4.8 & 44.00 - 45.00 &   8 &   6.9    
 &  $(3.2^{+1.5}_{-1.1})\cdot 10^{-6}$ & (6)\\  
3.2 - 4.8 & 45.00 - 46.00 &   1 &   3.2    
 &  $(1.7^{+3.5}_{-1.4})\cdot 10^{-8}$ \\  
3.2 - 4.8 & 46.00 - 47.00 &   3 &   1.9    
 &  $(9.8^{+8.8}_{-5.3})\cdot 10^{-10}$ \\ 
3.2 - 4.8 & 47.00 - 48.00 &   0 &   0.6    
 &  $< 2.9\cdot 10^{-11}$ \\               \\
\hline
\end{tabular}
\end{center}
\label{tab:exlf}
Notes:
$^{\rm a} L_{\rm x} h_{70}^{-2}\,{\rm erg\,s^{-1}}$ in the 0.5--2 keV band.
$^{\rm b} h_{70}^3\,{\rm Mpc}^{-3}$; the quoted errors are 68\%
Poisson errors using approximations by Gehrels \cite{gehrels}
of the number of AGNs.
\end{table*}

\subsection{Evolution of the Space Density}
\label{sec:evol}
 
\begin{figure*}
\centering
\resizebox{\hsize}{!}{\includegraphics{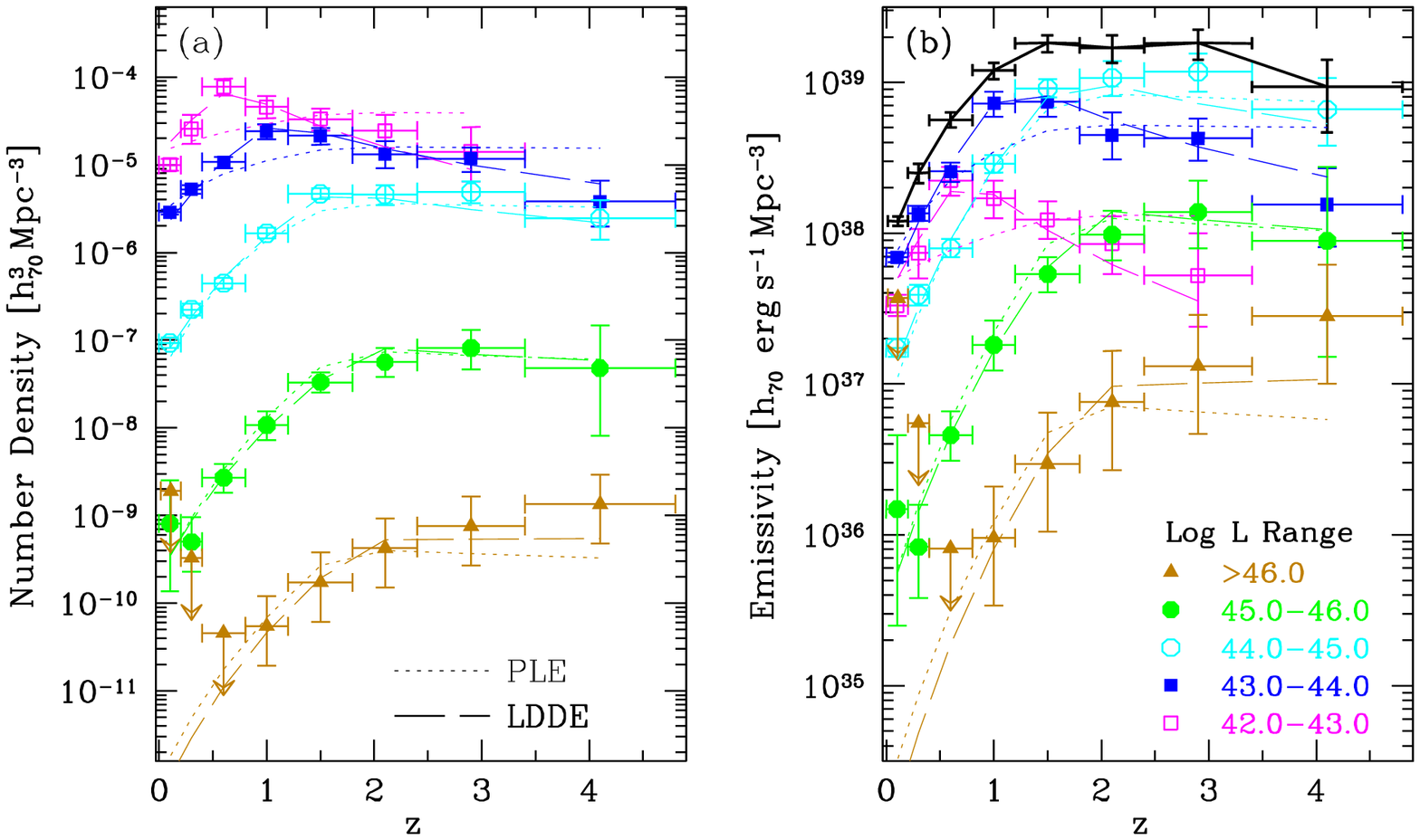}}
\caption{(a) The space density of AGNs as a function of redshift in 
different luminosity classes and the sum over all luminosities with 
${\rm log}\;L_{\rm x}\geq 42$. 
Densities from the PLE and LDDE models (Sect. \ref{sec:pleldde}) are 
overplotted with solid lines.
(b) The same as (a), except that the
soft X--ray emissivities are plotted instead of number densities. The
uppermost curve (black) shows the sum of emissivities in all luminosity
classes plotted.} 
\label{fig:ndens}
\end{figure*}

In this section, we investigate the evolution of the 
type--1 AGN space density in different luminosity classes as a 
function of redshift. The estimator of the space density
is the $N^{\rm obs}/N^{\rm mdl}$. The fit with Eq.~\ref{eq:2po}
has been made in finer redshift shells than in Sect.~\ref{sec:xlf}. 
The space densities as a function of redshift were calculated in five luminosity classes
with  ${\rm log}\;L_{\rm x}$ of 42--43, 43--44, 44--45, 45--46, and $>$46
as well as the sum over all luminosities with 
${\rm log}\;L_{\rm x}>42$. The resulting curves are shown in 
Fig.~\ref{fig:ndens}(a) for the nominal calculations. The incompleteness 
upper bounds have also been calculated, but have not been shown here for 
the visibility of the figure. These upper bounds are shown in Sect. \ref{sec:disc-comp}
(Figs. \ref{fig:comp_ma_ta} \& \ref{fig:zevcomp}). 
Since
the Black Hole growth function is more closely linked to the emissivity per 
comoving volume, we also show the emissivity as a function of redshift in the 
same luminosity classes in Fig.~\ref{fig:ndens}(b). 

Figure~\ref{fig:ndens}(a) clearly shows a shift of the number density peak 
with luminosity, in the sense that more luminous
AGNs (QSOs) peak earlier in the history of the universe, while the 
low luminosity ones arise later. Also, there is a clear decline of 
the derived space densities at least for luminosities of 
${\rm log}\;L_{\rm x} < 44$, even when the optical incompleteness 
upper bounds are taken into account. The counting statistics and 
spectroscopic incompleteness for the more luminous
AGNs do not allow to determine a decline, but do also 
not exclude it. This issue is further discussed in Sections~\ref{sec:mxs}
and \ref{sec:disc-comp}. 

In order to show the behaviour of the luminosity dependence of the
evolution more quantitatively, we have also made a maximum-likelihood fit
of the evolution curve in each of the luminosity bins, with $\log L_{\rm x}$
ranges of 42--43,43--44,44--45, and 45--46. We used two power law
components of the $(1+z)$ evolution with a cutoff redshift:
 
\begin{equation}
e_{\rm d}(z,L_{\rm xc}) = \left\{ 
	\begin{array}{ll}
	(1+z)^{p1} & (z \leq z_{\rm c}) \\ 
        e_{\rm d}(z_{\rm c})[(1+z)/(1+z_{\rm c})]^{p2} & (z > z_{\rm c})\\
	\end{array}
       \right. .
\label{eq:ev_d}
\end{equation}
where $L_{\rm xc}$ is the central (logarithmic) luminosity of the bin. 
As was the case for Eq. \ref{eq:2po}, the fit depends on the shape of 
the luminosity function (along the luminosity direction) within the luminosity
bin. Again, we have fixed the behavior in the luminosity
direction using those from the  LDDE model (Sect. \ref{sec:pleldde}) 
as a template.  We also show the normalization:  
\begin{equation}
A_0\equiv {\rm d}\Phi (L_{\rm x}=L_{\rm xc},z=0)/{\rm d}{\log L_{\rm x}}.
\end{equation}
The best-fit results are shown in Table~\ref{tab:zdevpars} together 
with the K--S probabilities. 
In the $\log L_{\rm x}=45-46$ bin, the fit for $p_2$ was unconstrained.  
Thus we have fixed the values of $p_2$ to that from LDDE (see below) for this
luminosity bin. 

\begin{table*}
\caption[]{Best--fit evolution parameters for each luminosity bin}
\begin{center}
\begin{tabular}{cccccccc}
\hline\hline
$\log L_{\rm x}$-range & $\log L_{xc}$ & $N$ & $A_0$
  & $p_1$ & $z_{\rm c}$ & $p_2$ & KS--prob$^{\rm b}$ \\
\hline                                               
42.0-43.0 & 42.5 & 117 &$(7.67\pm1.28)\cdot 10^{-6}$ &
 4.90$^{+1.21}_{-1.12}$ &  0.65$^{+0.12}_{-0.12}$ &
 -2.4$^{+1.0}_{-1.1}$ & 0.47,0.77,0.64\\
43.0-44.0 & 43.5 & 381 &$(1.59\pm0.15)\cdot 10^{-6}$ &
 3.89$^{+0.43}_{-0.50}$ &  1.11$^{+0.22}_{-0.11}$ &
 -1.8$^{+0.7}_{-1.1}$ & 0.55,0.25,0.39\\
44.0-45.0 & 44.5 & 303 &$(1.83\pm0.19)\cdot 10^{-8}$ &
 5.51$^{+0.38}_{-0.37}$ &  1.78$^{+0.14}_{-0.16}$ &
 -1.8$^{+1.3}_{-1.4}$ & 0.05,0.47,0.07\\
45.0-46.0 & 45.5 &  53 &$(4.90\pm1.21)\cdot 10^{-11}$ &
 6.06$^{+1.18}_{-1.22}$ &  1.79$^{+0.59}_{-0.26}$ &
 -0.4(*) & 0.81,0.98,0.62\\
\hline
\end{tabular}
\end{center}
\label{tab:zdevpars}
Parameter values which have been fixed during the fit are labelled
by '(*)'.  $^{\rm a}$Units -- $A_0$: $h_{70}^3\;{\rm Mpc^{-3}}$,\,\,
$L_{\rm x,*}$: $10^{44}\;h_{70}^{-2}{\rm erg\;s^{-1}}$.
$^{\rm b}$ The three values are probabilities in two
1D--KS test for the distribution, $L_{\rm x}$, 1D--KS test for the
$z$ distribution and the 2D--KS test for the ($L_{\rm x}$,$z$) space
respectively.
\end{table*}

\subsection{Global Representations by Analytical Functions}
\label{sec:pleldde}

It is sometimes useful to provide a simple analytical fit 
for the SXLF over the whole redshift-luminosity range. We
first used the pure-luminosity evolution (PLE) form, in order to 
enable a comparison with previous work:

\begin{equation}
\frac{{\rm d}\;\Phi\,(L_{\rm x},z)}{{\rm d\;log}\;L_{\rm x}}
  = \frac{{\rm d}\;\Phi\,(L_{\rm x}/e_{\rm l}(z),0)}
    {{\rm d\;log}\;L_{\rm x}},
\label{eq:ple}
\end{equation}

with the luminosity evolution factor:

\begin{equation}
e_{\rm l}(z) = \left\{ 
	\begin{array}{ll}
	(1+z)^{p1} & (z \leq z_{\rm c}) \\ 
        e(z_{\rm c})[(1+z)/(1+z_{\rm c})]^{p2} & (z > z_{\rm c})\\
	\end{array}
       \right. .
\label{eq:ez}
\end{equation}

We again used the smoothed two power law form (Eq.~\ref{eq:2po},
excluding the $z$--dependent factor) for the $z=0$ SXLF. 
The best--fit PLE parameters are shown in Table~\ref{tab:best} with 
the results of the K--S tests. The best--fit PLE model is overplotted 
with the binned SXLF in Figs.~\ref{fig:exlf} \& \ref{fig:ndens}
as dotted lines. It is apparent from the comparison in these figures,
especially the latter, PLE does not represent the behaviour of the 
low--luminosity ($\log L_{\rm x}\la 44$), intermediate redshift 
($0.5\la z \la 1.8$) regime, due to the rather restrictive nature of the PLE
form. 

As a more general analytical form for a refined representation
of the SXLF, we have explored the  luminosity-dependent density evolution
form (LDDE) form, originally suggested by Schmidt \& Green (\cite{sg83})
for describing optically-selected QSOs: 

\begin{equation}
\frac{{\rm d}\;\Phi\,(L_{\rm x},z)}{{\rm d\;log}\;L_{\rm x}}
  = \frac{{\rm d}\;\Phi\,(L_{\rm x},0)}
    {{\rm d\;log}\;L_{\rm x}}\cdot e_{\rm d}(z, L_{\rm x}),
\label{eq:ldde0}
\end{equation}

where $e_{\rm d}(z, L_{\rm x})$ is the density function normalized to z=0.
The results from Sect.~\ref{sec:evol} show that 
the peak number density shifts from $z\sim 0.7$ at 
${\rm log}\; L_{\rm x}\sim 42.5$ to $z>2$ at high luminosities. 
Based on a similar observation of a hard X--ray--selected sample, 
Ueda et al. \cite{ued03} used an expression where 
$z_{\rm c}$ is a simple function of  $L_{\rm x}$:

\begin{equation}
e_{\rm d}(z,L_{\rm x}) = \left\{ 
	\begin{array}{ll}
	(1+z)^{p1} & (z \leq z_{\rm c}) \\ 
        e_{\rm d}(z_{\rm c})[(1+z)/(1+z_{\rm c})]^{p2} & (z > z_{\rm c})\\
	\end{array}
       \right. .
\label{eq:ldde1}
\end{equation}

along with

\begin{equation}
z_{\rm c}(L_{\rm x}) = \left\{ 
	\begin{array}{ll}
	z_{\rm c,0}(L_{\rm x}/L_{\rm x,c})^\alpha & 
	(L_{\rm x} \leq L_{\rm x,c}) \\ 
        z_{\rm c,0} & (L_{\rm x}> L_{\rm x,c})\\
	\end{array}
       \right. .
\label{eq:ldde2}
\end{equation}

The results of the analysis in the previous section
shown in Table~\ref{tab:zdevpars} suggest that considering the
dependence of $p1$ and $p2$ on luminosity would still improve the 
fit. Thus we have also included the following for our full LDDE expression: 

\begin{eqnarray}
p1 (L_{\rm x})= p1_{44}+ \beta_1\;(\log L_{\rm x}-44)\\
p2 (L_{\rm x})= p2_{44}+ \beta_2\;(\log L_{\rm x}-44)
\end{eqnarray}

The best--fit parameters and the results of the K--S tests for the 
PLE and  LDDE models are summarized in Table~\ref{tab:best}. The best--fit
PLE and LDDE models are overplotted on Figs.~\ref{fig:exlf} and \ref{fig:ndens}
with dotted and dashed lines respectively. A detailed discussion of the 
comparison of model and data is given in Sect. \ref{sec:disc-comp}.

\begin{table*}
\caption[]{Best--fit parameters for global expressions}
\begin{center}
\begin{tabular}{cccc}
\hline\hline
                           & PLE & LDDE \\
\hline
\multicolumn{3}{c}{z=0 SXLF Parameters}\\
\hline
$A_{44}$               & $1.85\pm .11\;10^{-7}$ &$2.62\pm .16\;10^{-7}$\\
$\log L_{\rm x,*}$         & $43.33\pm .09$         &$43.94\pm .11$\\
$\gamma_1$                 & $0.39\pm.09$           & $0.87\pm .10$\\
$\gamma_2$                 & $2.29\pm.09$           & $2.57\pm .16$\\
\hline
\multicolumn{3}{c}{Evolution Parameters}\\
\hline
$p1$,$p1_{44}$              & $2.67\pm .15$       & $4.7\pm .3$     \\
$z_{\rm c}$,$z_{\rm c,44}$  & $1.70(+.20;-.24)$   & $1.42\pm .11$   \\
$p2$,$p2_{44}$              & $-0.2\pm .7$        & $-1.5\pm .7$    \\
$\alpha$                    & \ldots              & $0.21\pm 0.04$  \\
${\rm log } L_{\rm x,c}$    & \ldots              & $44.67*$        \\
$\beta_1$                   & \ldots              & $0.7\pm 0.3$    \\
$\beta_2$                   & \ldots              & $0.6\pm 0.8$     \\
\hline
\multicolumn{3}{c}{K--S Probabilities ($L_{\rm x},z,2D)$}\\
\hline
$0.015<z<4.8,{\rm log}\;L_{\rm x}>42$  & $0.10,0.10,0.05$ & $0.86,0.65,0.36$  \\
\hline
\end{tabular}
\end{center}
\label{tab:best}
 Parameter values which have been fixed during the fit are labeled
by '(*)'.  $^{\rm a}$Units -- A: $h_{70}^3\;{\rm Mpc^{-3}}$,\,\,
 $L_{\rm x,*}$: $10^{44}\;h_{70}^{-2}{\rm erg\;s^{-1}}$.
$^{\rm b}$ The three values are probabilities in two
1D--KS test for the distribution, $L_{\rm x}$, 1D--KS test for the
$z$ distribution and the 2D--KS test for the ($L_{\rm x}$,$z$) space
respectively.
\end{table*}

\section{An alternate approach using the $V_{\rm max}$ method}
\label{sec:mxs}

As described in the Introduction, the luminosity function derived 
from survey data binned in luminosity and redshift does not 
necessarily apply to the centers of the $(L_{\rm x},z)$ bins.
This binning bias tends to be especially a problem if data are scarce 
(often at higher redshifts) and gradients across bins are large.
The previous section describes a procedure that corrects the
binned space densities to first order. 

In this section, we avoid deriving densities from binned survey data. 
Instead, we use the $V_{\rm max}$ values of individual RBS sources to 
derive the zero redshift luminosity function. We then derive by iteration an
analytical density template at various $L_{\rm x}$ values that, together with the
zero redshift luminosity function, accounts for the observed
counts and redshifts of the deeper surveys. The end result of the
procedure is a set of observed values of the luminosity function
that apply to the centers of the $(L_{\rm x},z)$ bins, 
and that is quite insensitive to the precise template employed. 
A further advantage of employing $V_{\rm max}$ of individual sources is that 
it can be derived for two or more selection variables. This allows us to 
account for the effect of a spectroscopic magnitude limit in some of the 
deeper surveys beyond which the redshift is unknown for most of the sources. 
In the first use of $V_{\rm max}$, this feature was used to derive the 
luminosity function of radio quasars from a sample in which only the 
optically brightest objects had redshifts (Schmidt \cite{schmidt68}).

\subsection{Using $V_{\rm max}$ to derive the luminosity function}
\label{sec:vmax}

The derivation of a luminosity function from objects in a well defined
sample usually involves binning the observations in redshift and 
luminosity. If we make the bins in luminosity so small that each 
contains only one or zero objects then the luminosity function is 
composed of contributions from each of the individual sample objects.
In the limit, each of the objects contributes to the 
luminosity function a delta function of amplitude $1/V_{\rm max}$ 
at the object's luminosity $L$, where $V_{\rm max}$ is the co--moving, 
density--weighted volume over which the object can be observed 
within the sample limits in flux and solid angle. This luminosity
function will reproduce the source counts of the input sample exactly.

We write the luminosity function as

\begin{equation}
  $$ \Phi(L_{\rm x},z) = \Phi(L_{\rm x},0) \rho(z,L_{\rm x}) $$
\label{eq:lufu}
\end{equation}

where $\rho(z,L_{\rm x})$ is the space density,
or density evolution, normalized to z=0.
We approximate $\rho(z,L_{\rm x})$ by an analytical density template. For 
the redshift dependence we use at low redshifts a power law of $(1+z)$; 
at higher redshifts we adopt the shape of the density function used 
by Schmidt et al. (\cite{ssg95}) for optically selected quasars:

\begin{eqnarray}
 & & \rho_{\rm tem}(z)=(1+z)^m \;\;\;\;\;\;\;\;\;\;\;\;\;\;\;\;\;\;\;\; 0<z<z_c\\
\label{eq:tem1}
 & & \rho_{\rm tem}(z)=(1+z_c)^m \;\;\;\;\;\;\;\;\;\;\;\;\;\;\;\;\;\; z_c<z<z_d\\
\label{eq:tem2}
 & & \rho_{\rm tem}(z) = (1+z_c)^m 10^{k(z-z_d)} \; \; \; \; z > z_d
\label{eq:tem3}
\end{eqnarray}  

We use the RBS which contains 205 AGN--1 in the 
$\log L_{\rm x}$ range 42--46, mostly at low redshifts, to derive the
zero redshift luminosity function $\Phi(L_{\rm x},0)$.
The main excercise then is to derive the values of the template
parameters $(m,z_c,z_d,k)$ by fitting to the flux and redshift 
distributions in the deeper X--ray surveys. This will allow the
direct derivation of the luminosity function at the center of
bins of luminosity and redshift, as outlined in the steps below.

\begin{figure}
  \centering
  \includegraphics[angle=-90,width=9cm]{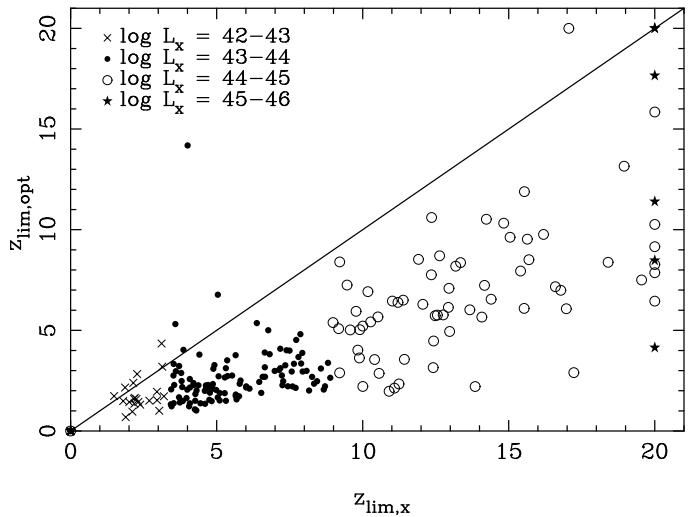}
  \caption{The X--ray redshift limit versus the optical redshift limit 
  for 205 RBS AGN--1 sources. In this illustration the X--ray flux limit
  is $10^{-16}$ cgs and the magnitude limit $R=24$, similar to 
  the case of the CDF--N.} 
  \label{fig:X_opt_limit_RBS}
\end{figure}

\begin{enumerate}
  \item Start the iteration by assuming initial values of the template
parameters $(m,z_c,z_d,k)$ as a function of $L_{\rm x}$; 

  \item The zero redshift luminosity function is the sum of delta 
functions for the assembly of RBS sources
\begin{equation}
 $$\Phi(L_{\rm x},0) = \sum_{i=1}^{205} (1/V_{{\rm max},i}^{\rm RBS}) \delta(L_{\rm i}-L_{\rm x})$$
\label{eq:step2}
\end{equation} 
where $V_{{\rm max},i}^{\rm RBS}$ is the accessible density--weighted volume of RBS 
source $i$ in the RBS, based on its solid angle and flux limit distribution;

  \item Next, predict expected numbers and redshifts in the deeper surveys. 
For the part of the luminosity function based on RBS source $i$, 
the expected number in survey $sur$ based on the density template is
\begin{equation}
$$n_{\rm tem}(L_{\rm i},z) \Delta z = 
(V_{{\rm max},i}^{sur}(z)/V_{{\rm max},i}^{\rm RBS}) \Delta z$$
\end{equation} 
where $V_{{\rm max},i}^{sur}(z)$ is the accessible density--weighted volume of 
RBS source $i$ over the redshift range $z,z+\Delta z$ 
based on the solid angle and flux limit distributions in survey $sur$;
 
  \item In order to compare with the observed numbers, we use
four luminosity classes with $\Delta \log L_{\rm x}=1.0$ centered on 
$\log L_{\rm x}=42.5, 43.5, 44.5, 45.5$ and redshift shells of
$\Delta \log z = 0.1$ centered on $ \log z = -0.95, -0.85, ..., 0.75$.
Total binned numbers predicted in all the non--RBS surveys are 
$N_{\rm tem}(L_{\rm x},z)$;

  \item Next, use the observed number $N_{\rm obs}(L_{\rm x},z)$ in each $(L_{\rm x},z)$ 
bin to derive the space density or luminosity function at the
center of each luminosity bin by scaling the template value of the 
luminosity function: 
\begin{equation}
$$\Phi_{\rm obs}(L_{\rm x},z)=\Phi(L_{\rm x},0)\rho_{\rm tem}(z,L_{\rm x})N_{\rm obs}(L_{\rm x},z)/N_{\rm tem}(L_{\rm x},z)$$
\label{eq:scale}
\end{equation}
 
  \item Comparison of $\Phi_{\rm obs}(z,L_{\rm x})$ and $\Phi(L_{\rm x},0)\rho_{\rm tem}(z,L_{\rm x})$ 
serves as a guide for the next iteration of the template parameters 
$(m,z_{\rm c},z_{\rm d},k)$, starting at step 2 above.
\end{enumerate}

\begin{figure}
  \centering
  \includegraphics[angle=0,width=9cm]{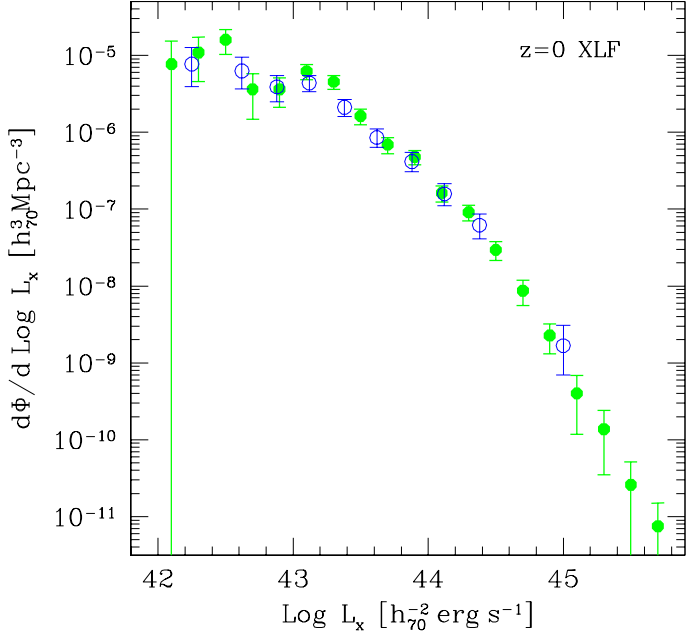}
  \caption{The z=0 luminosity function $\Phi(L_{\rm x},0)$
           derived from 205 RBS AGN--1 sources using the $V_{\rm max}$
           method (red filled symbols), compared with the 
           binned estimate in the redshift shell z=0.015--0.2,
           corrected to z=0.}
  \label{fig:lufu}
\end{figure}

\subsection{Deriving $V_{\rm max}$ values}
\label{sec:derivevmax}

The values of $V_{{\rm max},i}^{sur}$ are based on the maximum accessible redshifts
$z_{{\rm max},i}^{sur}$ of RBS source $i$ in survey $sur$. For the SA--N, NEPS, 
RIXOS, RMS and CDF--S surveys, where missing redshifts are not correlated
with X-ray or optical flux, we assume that $z_{{\rm max},i}^{sur}$ equals the X-ray
redshift limit $z_{\rm lim,x}$. In accounting for the missing redshifts, 
we assume that their distribution is similar to that of the observed 
redshifts and use an effective solid angle appropriately reduced from the 
geometric solid angle. 

For the RDS/XMM and CDF--N surveys, where redshifts are missing 
primarily for optically very faint objects, we proceed as follows.
In this case, we employ only objects brighter than a magnitude limit
$m_{\rm lim}$ (in practice $R<24$) and use the full geometric survey solid angle. 
In predicting the number of expected objects in survey $sur$,
this introduces a second limiting redshift, $z_{{\rm lim,opt},i}$ that will
depend on the optical luminosity of RBS source $i$ contributing
to the luminosity function. The relevant limit $z_{{\rm max},i}^{sur}$ 
for the derivation of the expected counts is the smaller of
$z_{{\rm lim,x},i}$ and $z_{{\rm lim,opt},i}$.

The determination of $z_{{\rm lim,opt},i}$ is based on the magnitudes of the
RBS objects. Among the 205 RBS AGN--1, Salvato (\cite{salvato}) has
carried out optical photometry of a redshift-limited sample of 89 
sources, deriving magnitudes for the nucleus, disk and spheroid. 
Comparison of the total magnitudes given by Salvato with the RBS 
magnitudes shows $<R_{\rm RBS}- R_{\rm Sal}> = -0.5$ with a dispersion of 
0.9 mag. We apply this correction to the 116 remaining RBS sources. 
Based on the systematics of the Salvato magnitudes, we find that on 
the average, the absolute R--magnitude of the galaxy light is --21.3, 
that the disk light accounts for 56\% of the galaxy light, and that for 
low--luminosity objects the nucleus contributes no less that 25\%
of the total light. This allows the, admittedly uncertain, derivation
of the nuclear, disk and spheroid magnitudes of the remaining 116 RBS
objects. We use the following values for the spectral index
$\alpha = d \log S/d \log \nu$ of the nuclear, disk and spheroid 
components, estimated from spectral energy distributions given by 
Kinney (\cite{kinney}): $\alpha_{\rm nuc} = -0.5$, $ \alpha_{\rm dis} = -1.0$,
$\alpha_{\rm sph} = -3.0$. These values allow evaluation of the K-correction
required to derive the redshift $z_{{\rm lim,opt},i}$ at which each of the
205 RBS sources reaches the spectroscopic magnitude limit $R_{\rm lim}=24$.
We illustrate in Figure~\ref{fig:X_opt_limit_RBS} the effect of the optical 
limit on the redshift limits for the CDF--N: for objects below the diagonal 
line, the redshift limit is that set by the optical limit.

\subsection{The zero redshift X--ray luminosity function}
\label{sec:zerolufu}

The zero redshift luminosity function $\Phi(L_{\rm x},0)$ is derived from the 
RBS sources as outlined in Sect~\ref{sec:vmax}, Eq.~\ref{eq:step2}.
Since the accessible volume $V_{{\rm max},i}^{\rm RBS}$ is density-weighted, 
this derivation requires information
about the density template, discussed in the next section.
At the low redshifts of the RBS objects (typically 0.1), this
involves only the $m$ parameter of the low redshift $(1+z)^m$ density 
variation. A binned version of $\Phi(L_{\rm x},0)$, created by summing the 
delta functions in bins of $\Delta \log L_{\rm x} = 0.2$, is shown in 
Fig.~\ref{fig:lufu}. 

\begin{figure*}
  \includegraphics[angle=-90,width=8truecm]{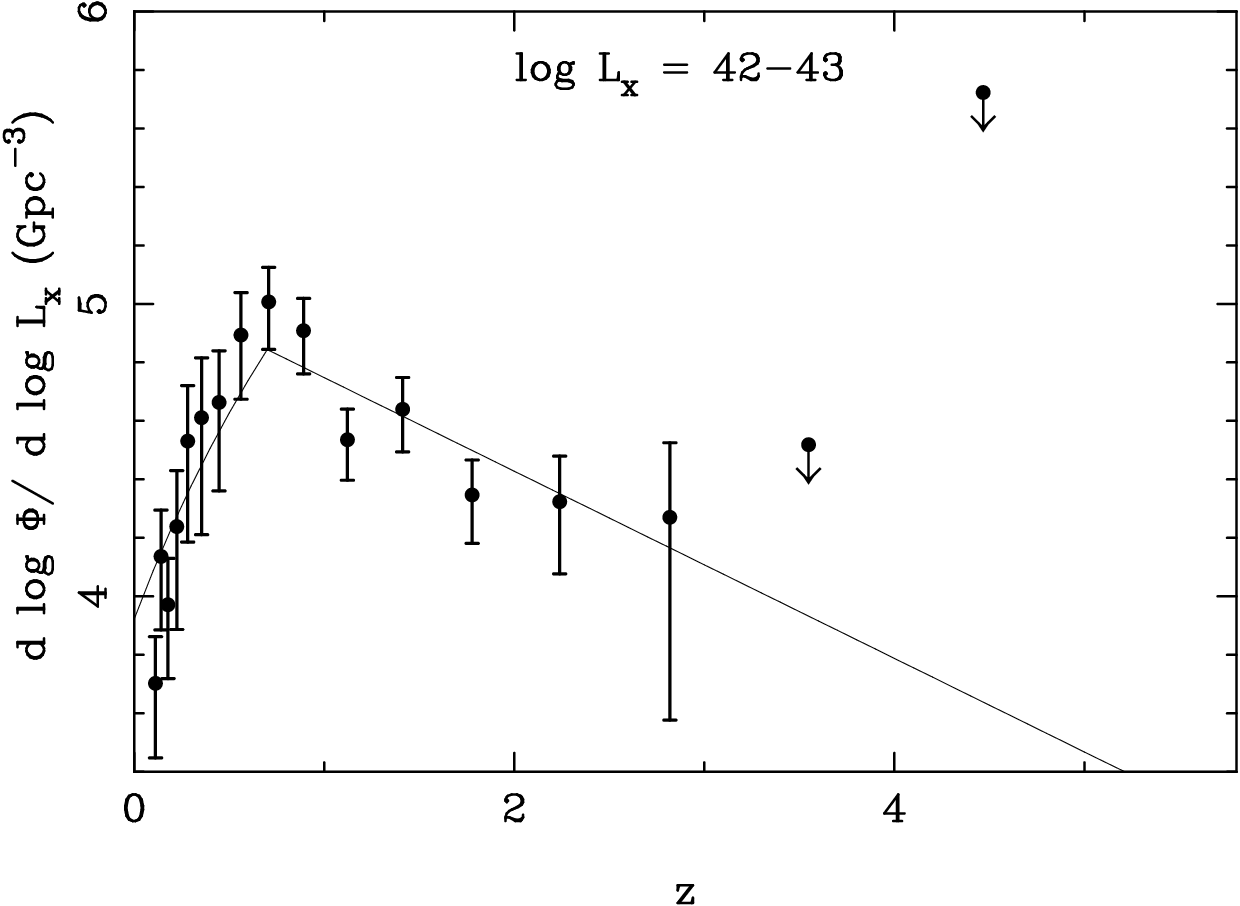}
  \includegraphics[angle=-90,width=8truecm]{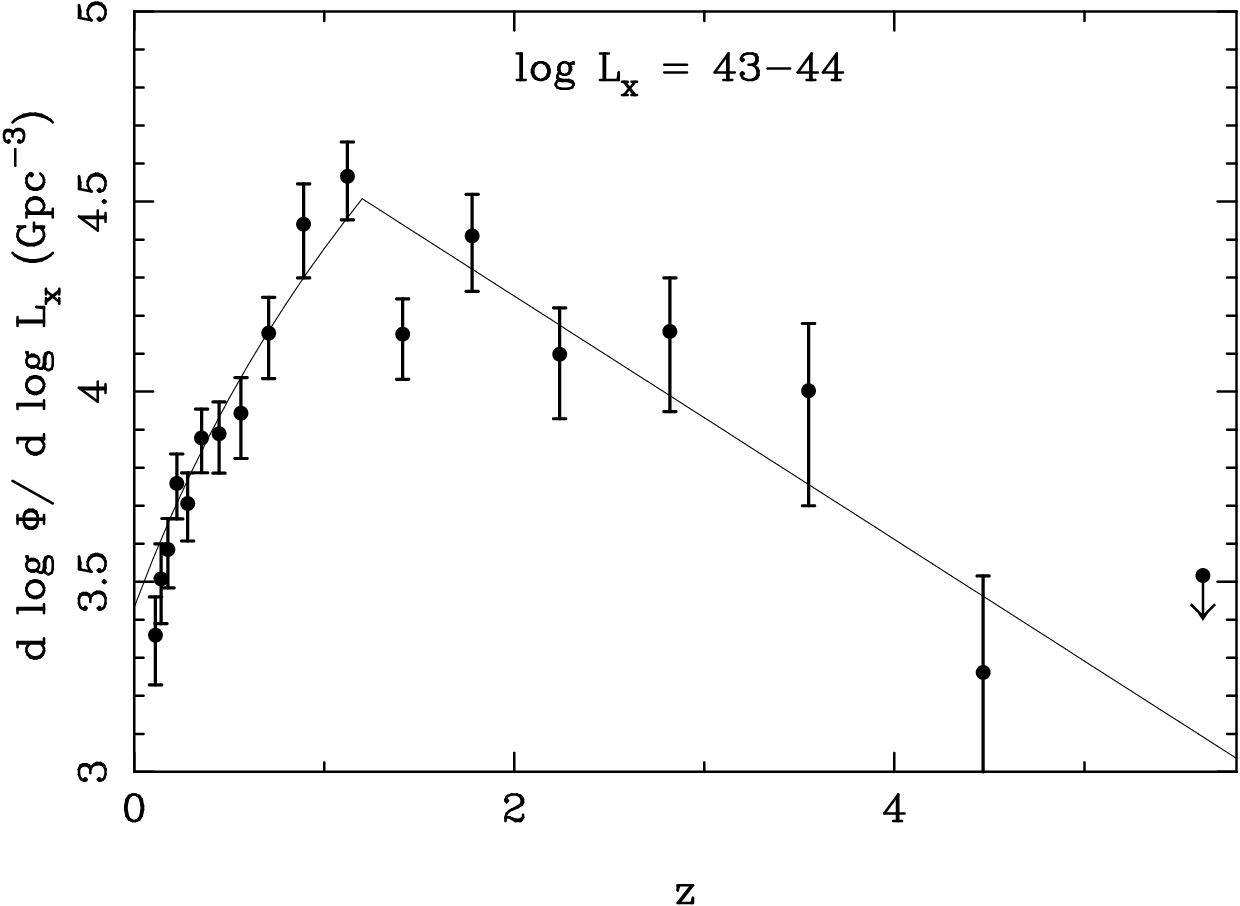}
  \caption{The luminosity function as a function of redshift $z$.
   The $z=0$ values are based on 21 RBS sources for $\log L_{\rm x}=42-43$, 
   111 sources for $\log L_{\rm x}=43-44$.   
   The points are based on numbers observed in the non--RBS surveys.
   The thin line represents the density template (see text for details).} 
  \label{fig:den4243}
\end{figure*}
\begin{figure*}   
  \includegraphics[angle=-90,width=8truecm]{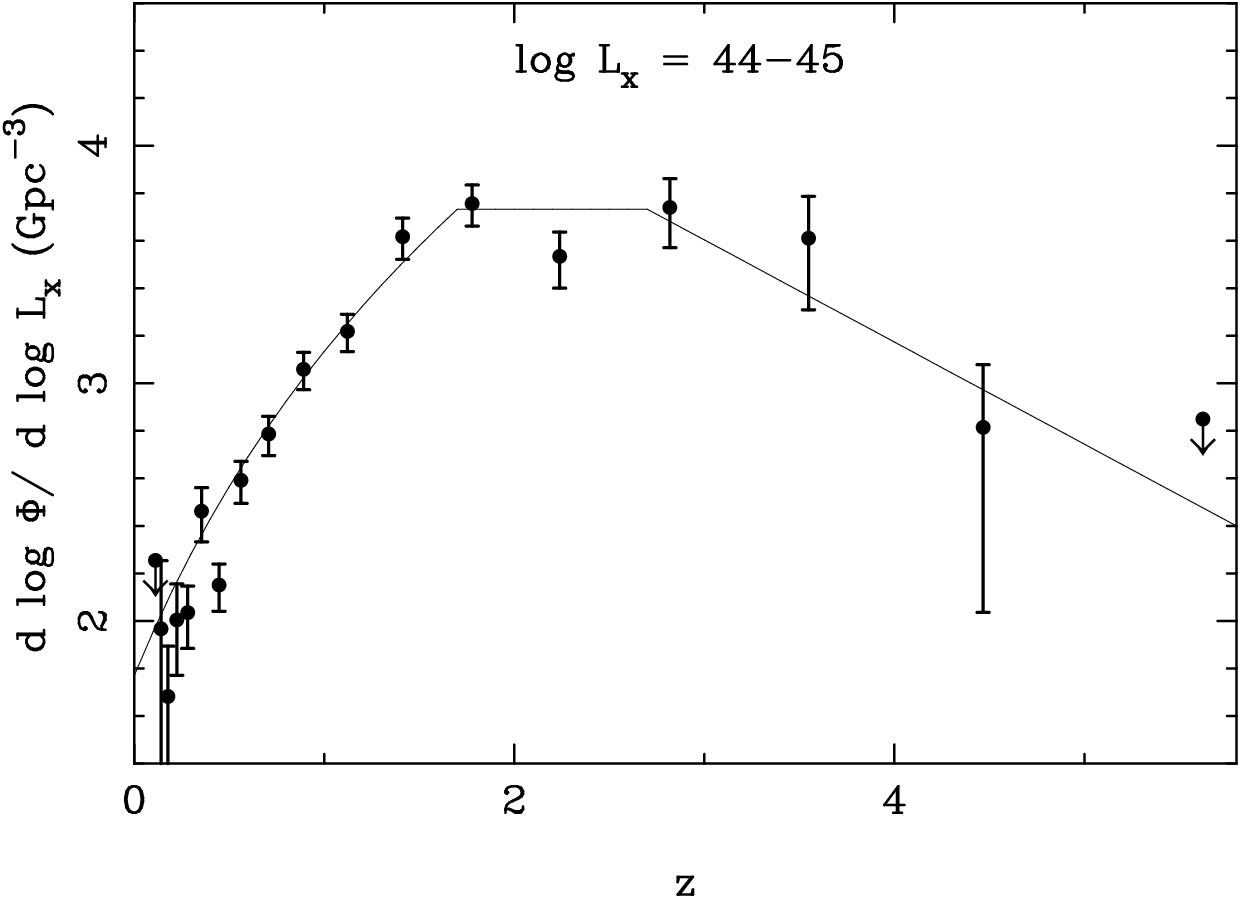}
  \includegraphics[angle=-90,width=8truecm]{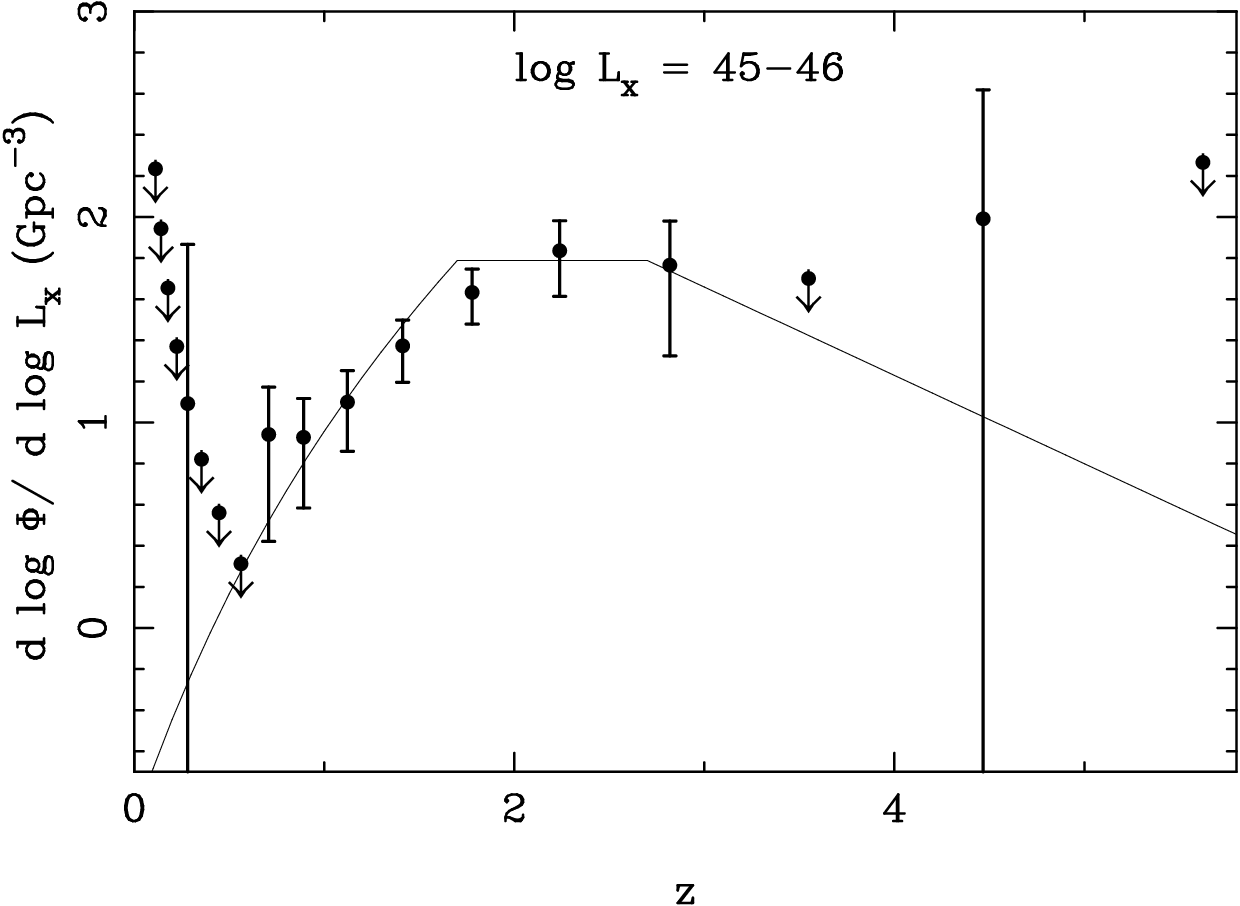}
  \caption{The luminosity function as a function of redshift $z$. 
   The $z=0$ value are based on 67 RBS sources for $\log L_{\rm x}=44-45$,   
   6 sources for $\log L_{\rm x}=45-46$.   
   The points are based on numbers observed in the non--RBS surveys.
   The thin line represents the density template; the parameters of
   the template for $z>1.7$ were adopted from those found for 
   high-luminosity optical quasars (see text for details).} 
  \label{fig:den4445}
\end{figure*}

\subsection{The analytical density template $\rho(z,L_{\rm x})$}

\label{sec:tem}

The density template defined in Eqs.~\ref{eq:tem1}--\ref{eq:tem3}
with parameters $(m,z_{\rm c},z_{\rm d},k)$
is an analytical approximation of the density function $\rho(z,L_{\rm x})$.
The derivation of the template parameters was carried out as follows. As 
described in Sect.~\ref{sec:vmax}, it is a procedure of trial and error. 
For each of the four luminosity classes, in iterating the value of 
$m$, we minimize $<N_{\rm obs}(L_{\rm x},z)-N_{\rm tem}(L_{\rm x},z)>$ for $0.5<z<z_{\rm c}$, 
since at low redshifts the luminosity function is firmly anchored by 
the RBS objects. The values of $z_{\rm c}$, $z_{\rm d}$ and $k$ are derived by
minimizing $<N_{\rm obs}(L_{\rm x},z)-N_{\rm tem}(L_{\rm x},z)>$ for $z>z_{\rm c}$, by adjusting
$k$ for the lower luminosity classes to fit the total observed AGN--1 
in the CDF--S and CDF--N (regardless of availability of redshifts or 
identification), and for the higher luminosity classes by adopting
parameter values
for optically selected quasars (see next section). This 
procedure makes it difficult to evaluate the statistical significance
of the templates. We will show below that the accuracy of the templates
actually has a negligible effect on the values of $\Phi_{\rm obs}(L_{\rm x},z)$.

\begin{table}
  \caption[]{Parameters characterizing the density template.} 
     \label{tab:tem}
 $$
     \begin{array}{lcccc}
        \hline
        \noalign{\smallskip}
        \log L_{\rm x}                    &  42.5  &  43.5  &  44.5  &  45.5 \\
        \hline
        \noalign{\smallskip}
         m     & 4.0 \pm 0.7 & 3.4 \pm 0.5 & 5.0 \pm 0.2 & 7.1 \pm 1.0 \\  
         z_{\rm c}                        &   0.7  &   1.2  &   1.7  &   1.7 \\  
         z_{\rm d}                        &   0.7  &   1.2  &   2.7  &   2.7 \\  
         k                          &  -0.32 &  -0.32 &  -0.43 &  -0.43\\  
        \noalign{\smallskip}
        \hline
     \end{array}
 $$
\end{table}

The resulting values of $(m,z_{\rm c},z_{\rm d},k)$ for the four luminosity classes are 
given in Table~\ref{tab:tem}. The errors on for the low--redshift 
power index $m = d \log \rho/d \log (1+z)$ have been estimated using 
the number of observed objects in the redshift shells $0.5<z<z_{\rm c}$ in
comparison with the RBS objects. From this analysis and the results shown in 
Table~\ref{tab:zdevpars} it appears that $m$ exhibits a
significant increase with $L_{\rm x}$ for $\log L_{\rm x} > 43.0$. For these
luminosities, we employ a quadratic interpolation of $m$ with $\log L_{\rm x}$ 
through the values given in the table. For $\log L_{\rm x} = 42-43$ we assume 
$m$ to be constant. 

\begin{figure}
  \centering
  \includegraphics[angle=-90,width=9cm]{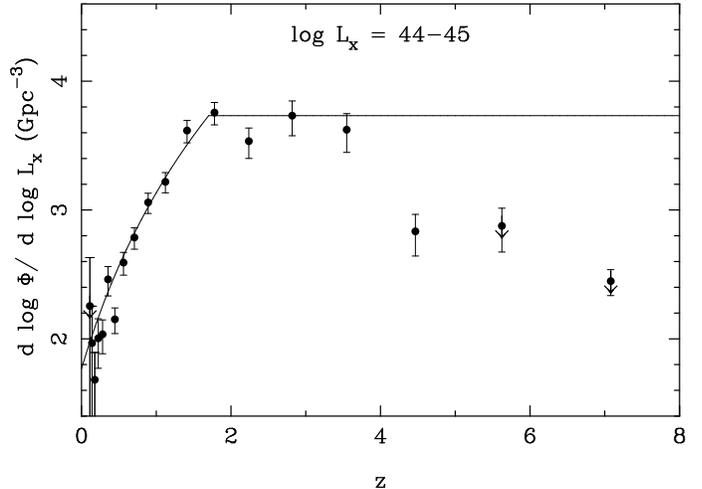}
  \caption{The luminosity function versus redshift $z$, derived using
   a template having constant space density at large redshift. 
   The observed numbers at $z = 4.47, 5.62, 7.08$ are 1, 0, 0, 
   respectively. This plot gives clear evidence that high-luminosity
   X-ray AGNs type 1 show a decline in space density beyond $z \sim 4$.}
  \label{fig:den44flat}
\end{figure}

\subsection{The X--ray luminosity function $\Phi_{\rm obs}(L_{\rm x},z)$}
\label{sec:lufu}

The observed values of the luminosity function $\Phi_{\rm obs}(L_{\rm x},z)$
are obtained by scaling the template luminosity function by a factor
$N_{\rm obs}(L_{\rm x},z)/N_{\rm tem}(L_{\rm x},z)$, see Eq.~\ref{eq:scale}. These values 
of $\Phi_{\rm obs}(L_{\rm x},z)$ are plotted versus the redshift for each of the 
four luminosity classes in Figs.~\ref{fig:den4243}--\ref{fig:den4445}. 
The $\pm 1\sigma$ error bars 
are based on the numbers $N_{\rm tem}(L_{\rm x},z)$ predicted by the template.
The figures also show the template luminosity function resulting from 
the product of the zero redshift luminosity function $\Phi(L_{\rm x},0)$ 
and the template density function $\rho_{\rm tem}(z,L_{\rm x})$.

The $\pm 1\sigma$ error bars do not include any contribution reflecting 
the error of the template density function $\rho_{\rm tem}(z,L_{\rm x})$. We
explore the effect of the template by rederiving the predicted number
$N_{\rm tem}(L_{\rm x},z)$ on the extreme assumption that $m=0$ and $k=0$, i.e
that there is zero density evolution. We find that the observed
densities $\log \Phi_{\rm obs}(L_{\rm x},z)$ increase by only $0.00-0.06$. 
Since the template errors must be much smaller than assumed in this 
extreme example, their effect on the error of $\Phi_{\rm obs}(L_{\rm x},z)$ 
is negligible. {\sl The
main result of this section is the set of $\Phi_{\rm obs}(L_{\rm x},z)$ values
plotted in Figs.~\ref{fig:den4243}--\ref{fig:den4445}; the templates
serve primarily to eliminate the uncertainty related to binning.} 

For $\log L_{\rm x} = 42-43$ and $43-44$, the density rises by an order of
magnitude to $z \sim 0.7$ and $1.2$, respectively, and then declines
steadily (Fig.~\ref{fig:den4243}). As shown in the next section, the
peak at $z \sim 0.7$ for $\log L_{\rm x} = 42-43$ is little affected by
redshift spikes caused by large-scale structure.

The density distribution for AGN--1 with $\log L_{\rm x}$=44--45 is documented 
to $z \sim 4$ (Fig.~\ref{fig:den4445}). Since in this case, the evidence 
for a decline in density at high redshift was not initially clear, we
adopted for the parameters $(z_{\rm c},z_{\rm d},k)$ of the template above $z=1.7$ 
those found for high--luminosity optical quasars by Schmidt et al. 
(\cite{ssg95}). It appears from Fig.~\ref{fig:den4445} that the X--ray data 
are consistent with the adopted shape. In order to further investigate
whether the space density declines significantly at high redshift,
we explore a test template in which the density 
does not decline at all, i.e. remains flat at high redshifts
$(k=0)$. The results are shown in Fig.~\ref{fig:den44flat}. The error 
bars are much reduced from those in Fig.~\ref{fig:den4445}, reflecting the
fact that our errors are based on the {\it predicted} numbers, which
are larger for a flat evolution function at high redshift. The bins at 
$z = 4.47, 5.62, 7.07$ have observed/predicted numbers of 1/7.9, 0/7.2,
and 0/19.3, respectively. Limiting the case to $z<6.3$, for which
the observation of redshifts in high-luminosity quasars should be no
problem, we have for $z>4$ one observed object versus 15 expected.
The Poisson probability for such an occurence barring systematic effects
is $3 \times 10^{-7}$, constituting strong evidence that for 
$\log L_{\rm x}=44-45$ the space density declines beyond $z \sim 4$.

The low-redshift density parameter $m$ for $\log L_{\rm x} = 45-46$ 
continues the trend of an increasing $m$ for larger $L_{\rm x}$. At large
redshift, the sparse data do not give any information about the density 
beyond $z \sim 3$ (see Fig.~\ref{fig:den4445}).

\section{Discussion}
\label{sec:disc-comp}

In the present paper we have used two different methods to derive the 
AGN--1 X--ray luminosity function and its evolution. Detailed 
descriptions of the two methods are given in Sects.~\ref{sec:exlf1}
and \ref{sec:vmax}. Conceptually, the binned method derives a first
order luminosity function by dividing the numbers $N_{\rm obs}(L_{\rm x},z)$ 
observed in the input surveys by the appropriate volumes. An 
analytical representation of the luminosity function is used 
to predict the numbers $N_{\rm mdl}(L_{\rm x},z)$ expected in these surveys. The 
luminosity function is then corrected by the factor $N_{\rm obs}/N_{\rm mdl}$. 

In the $V_{\rm max}$ method, the RBS is used to derive the zero redshift
luminosity function. An analytical density template is 
used to predict the numbers $N_{\rm tem}(L_{\rm x},z)$ expected in the
deeper surveys. In this process, the effect of a spectroscopic magnitude 
limit on the $V_{\rm max}$ values is included for deep surveys where this
limit applies. Once the template predictions are close to the observed
numbers $N_{\rm obs}(L_{\rm x},z)$ in the deeper surveys, the luminosity function
at the center of each $(L_{\rm x},z)$ bin is derived by multiplying the
template luminosity function by $N_{\rm obs}/N_{\rm tem}$. 

Using the optical magnitudes and spectra of the RBS sources in 
deriving the redshifts at which they would reach the spectroscopic limit
$R=24$ introduces uncertainties. The magnitudes of the 116 RBS sources 
not studied by Salvato (\cite{salvato}) are quite poor and so is their
separation in nucleus, disk and spheroid components. The assumed 
spectra of these components are schematic. Obtaining spectral energy
distributions for all the RBS sources down to the far UV would allow
deriving their flux directly at or near the rest wavelengths corresponding 
to the $R$ magnitude at the limiting redshifts.  

\begin{figure*}
  \centering
  \includegraphics[width=15cm]{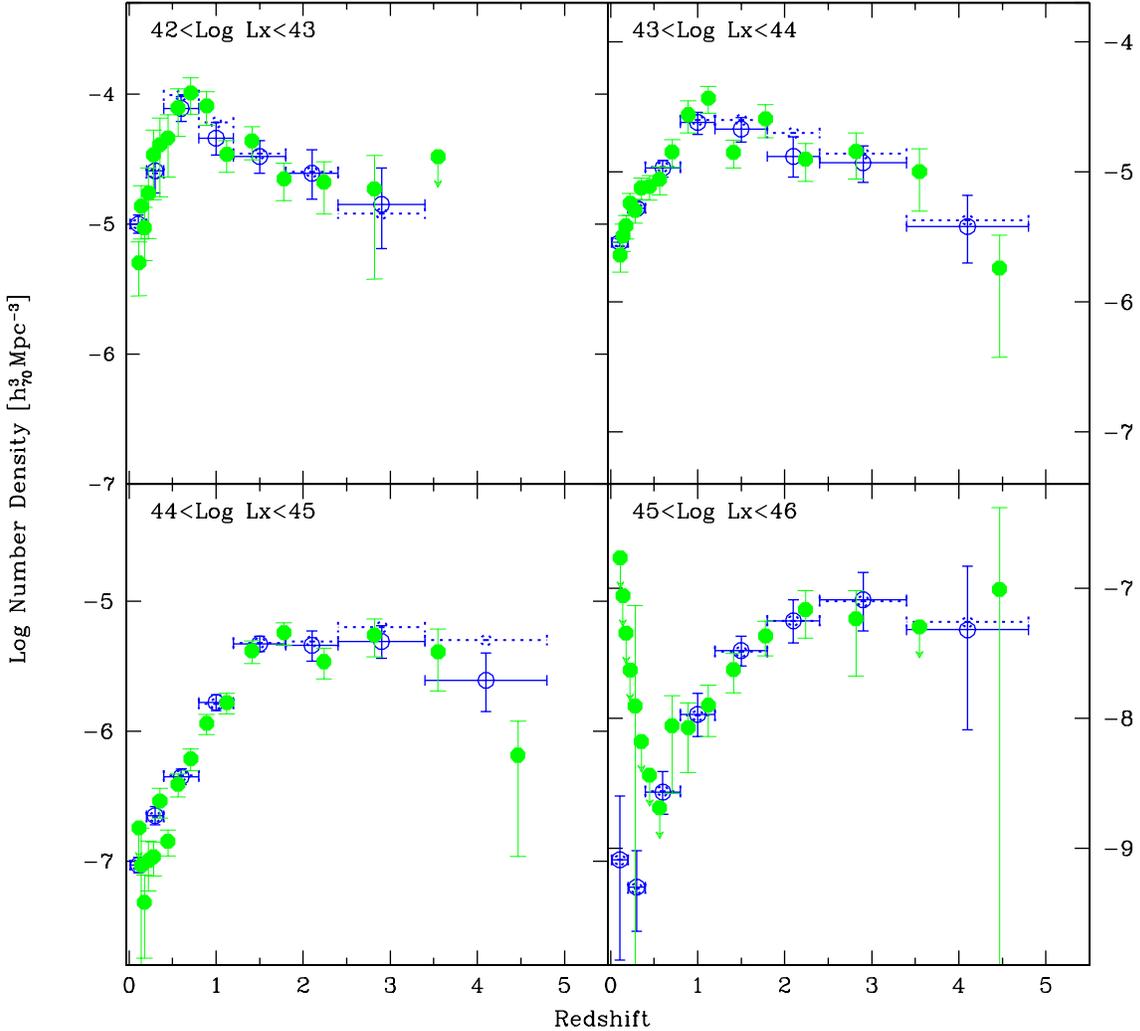}
  \caption{Comparison between the space densities derived  
           with two different methods. The blue datapoints with error 
           bars refer to the binned treatment using the 
           N$_{\rm obs}$/N$_{\rm mdl}$ method, the dashed horizontal 
           lines corresponding to the maximum contribution of unidentified
           sources. The red dots refer to the $V_{\rm max}$ method and the 
           data given in Figs.~\ref{fig:den4243} and~\ref{fig:den4445}.}
  \label{fig:comp_ma_ta}
\end{figure*}

It is reassuring that the general properties and absolute values of the 
space density are very similar in the two different derivations in 
Sections~\ref{sec:exlf} and \ref{sec:mxs}. Figure~\ref{fig:comp_ma_ta} 
shows a direct comparison between the binned and $V_{\rm max}$ determinations 
of the space density, which agree very well within statistical errors. 

We use overall fits to the luminosity function and its cosmological 
evolution in order to compare to previous work and to enable theoretical 
calculations with simplified analytical forms. For this purpose, we use 
two functional forms, i.e. a pure luminosity evolution (PLE) and 
a luminosity--dependent density evolution (LDDE) 
model. Assuming PLE, the luminosity evolution index $p1\approx 2.7$ and 
the cutoff redshift $z_{\rm c}\approx 1.7$ are in rough agreement with 
previous results from type--1 AGNs in {\it ROSAT} or {Einstein/ROSAT} 
combined surveys (e.g. paper I; Jones et al.\cite{jones97}; 
Page et al.\cite{page97}; Boyle et al. \cite{boy93}). 
    
The results of the K--S test for the PLE fit over the whole sample are 
marginally acceptable, with $\sim 5\%$ chance of obtaining a 
2D K--S value larger than observed. However, this is caused by the sheer
number of AGNs in the part of the z--L$_{\rm x}$ space where PLE is still 
a good description, which dominates the overall statistics. 
Figure~\ref{fig:ndens} clearly shows that PLE fails to reproduce the 
behavior of the SXLF around $\log L_{\rm x}\la 44$ at $0.4\la z \la 1.7$. 
The luminosity bin $42<{\rm log\,L_{\rm x}<43.5}$ in the 
$0.4<z<0.8$ shell contains 41 objects where PLE predicts 19 and the adjacent  
regime of $43.0<{\rm log\,L_{\rm x}<44.0}$ at $0.8<z<1.2$ contains 46 objects
where PLE predicts 25, so that the overall excess corresponds to more than $4.5 \sigma$. 
In constructing the LDDE form, we tried to fully represent 
the SXLF from our data with unprecedented redshift and luminosity
coverage. The overall 2D K--S acceptance of LDDE has been improved to 36\%. 
The only location where the LDDE model still deviates from the data significantly 
is the very end of our sample, the $42<{\rm log\;}<43$ bin at $0.015<z<0.2$, 
where LDDE predicts 66 objects while we observed 45 objects 
(a $\sim 3 \sigma$ deviation).

Even though our sample is a soft X--ray-selected type--1 AGN sample,
the overall behaviour of our XLF is similar to that obtained by Ueda et
al. (\cite{ued03}) for the intrinsic (de--absorbed) luminosity function of hard
X--ray selected obscured and unobscured AGN. To make this comparison,
we have refitted our sample with the LDDE model where $\beta_1$ and $\beta_2$
are fixed to zero. This is exactly the same as the function form which Ueda et al. (\cite{ued03}) used to describe their intrinsic HXLF. All the $z=0$ XLF parameters 
and evolution parameters are remarkably close between our SXLF and HXLF, except 
the global normalization. The HXLF normalization is found to be about five times 
larger than that of our SXLF, after adjustments for the differences in energy bands
and the difference in the luminosities at which normalizations are evaluated.  
This factor probably accounts for the absorbed objects missing in the SXLF.  
However, the Ueda et al. sample, containing about
250 AGN, is limited to lower luminosities and lower redshifts than our sample
of 1000 objects, so that its statistical quality and limited sensitivity range
were not sufficient to constrain the decline of the space density at high
redshifts, which has been measured significantly in our sample for the first
time.  

Very recently, Barger et al., \cite{bar05} have presented X--ray
luminosity function analyses both in the hard and soft X-ray bands, based on
the CDF--N, CDF--S, CLASXS and ASCA surveys. Again, their results are in good
agreement with the soft XLF discussed here and the hard XLF presented by Ueda
et al., however, they still suffer from substantial identification
incompleteness. Also, their results on broad-line AGN are not directly
comparable to our type--1 AGN sample, because they only include the optically
classified type--1 AGN and thus miss most of the low--luminosity unabsorbed AGN--1
we are concerned with in this paper. A critical comparison of our XLF with
those of Ueda et al., \cite{ued03} and Barger et al., \cite{bar05} will be the
topic of a future paper. 

The evidence for a peak in the evolution function is quite strong at
$z \sim 0.7$ for $L_{\rm x} = 42-43$ (see Figs.~\ref{fig:ndens}(a)
and ~\ref{fig:den4243}. 

The faintest end of the sample depends on the small field of
 Chandra Deep Field-South and there is some concern on the effects of
 the large-scale structure and cosmic variance associated with it.
 In particular, Gilli et al. \cite{gil03} found two
 redshift spikes, one  at $z\sim 0.67$ and the other at $z\sim 0.73$.
One may wonder whether the z=0.7 peak 
is caused by these redshift spikes (see also Gilli et al., \cite{gil05}).
 These do, however, not affect our SXLF estimates significantly. In the 
analysis in Sect.~\ref{sec:exlf} only six out
 of 41 AGNs (15\%) in our sample in the $0.4<z\leq 0.8$ bin in the range
 $42\leq {\rm log}\,L_{\rm x}<43.5$ are in these spikes
 ($0.664\leq z\leq 0.685$ and $0.725\leq z\leq 0.742$) and this is
 the only regime where the spikes give a non-negligible
 contribution. 
 In the analysis in Sect.~\ref{sec:mxs}, the z=0.71 bin contains 5 objects
 from the two spikes, for a total of 5 out of 15 observed objects.
 Disregarding these objects would decrease the derived density by dex 0.22 
 in Fig.~\ref{fig:den4243}, actually leading to better agreement with the
 template densities. We therefore conclude that cosmic variance is not
 significantly affecting our results on the evolution of the space
 density.

\begin{figure}
  \centering
  \includegraphics[angle=-90,width=9cm]{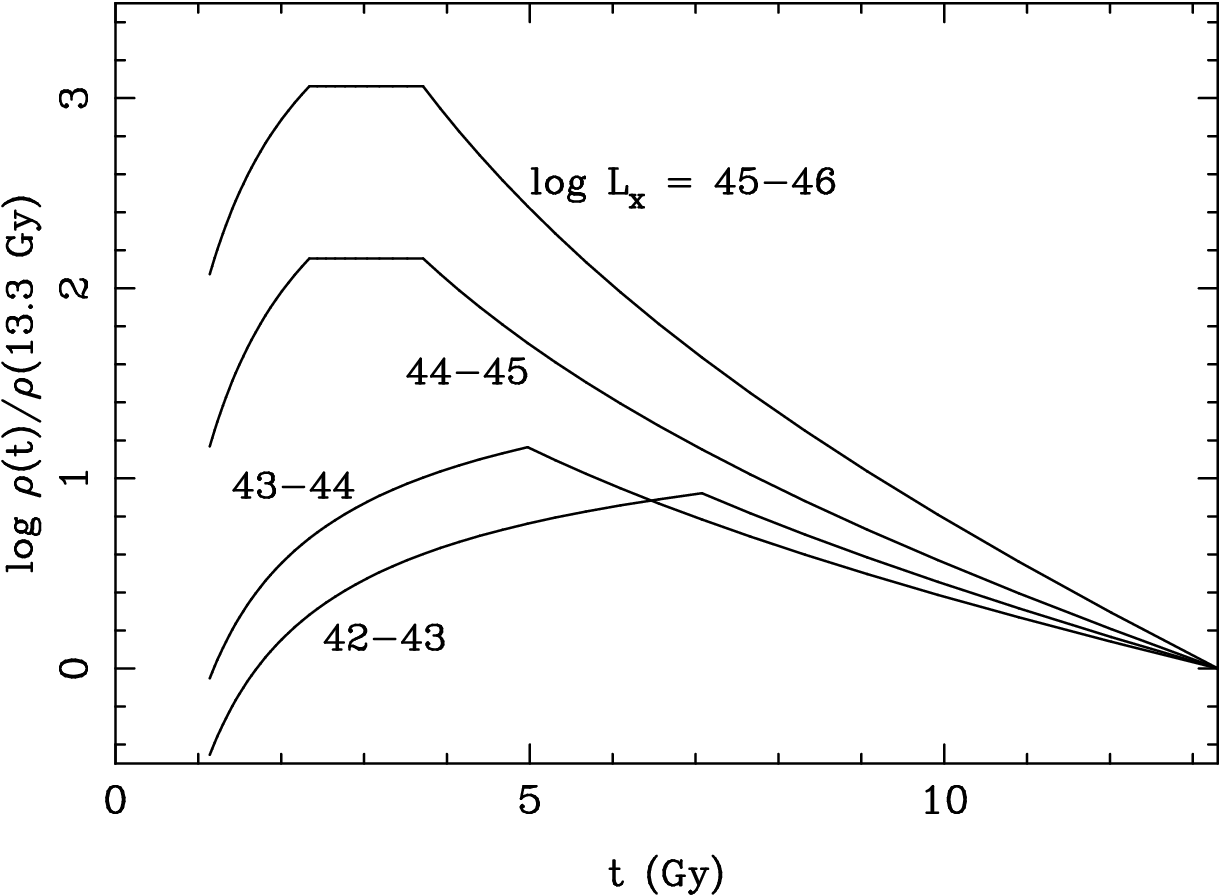}
  \caption{Co--moving density of AGN--1 versus cosmic time normalized 
           to the present time.}
  \label{fig:density_time}
\end{figure}

We show in Figure~\ref{fig:density_time} the AGN--1 space density as a 
function of cosmic time. 
We see dramatic changes with $L_{\rm x}$. For declining $L_{\rm x}$ as we move from
high-luminosity AGN or quasars to Seyfert galaxies, the main formation
of the objects occurred at later cosmic times. For $\log L_{\rm x}>44$ the density
curve is similar to that for quasars. It is an intriguing question
whether the observed dependence of $m$ on $L_{\rm x}$ is accompanied by a
corresponding dependence on $L_{\rm opt}$. At $\log L_{\rm x}<44$ the AGN--1 are
mostly Seyfert galaxies, for which there is no comparable optical evidence 
about their density curve. 

These new results paint a dramatically different 
evolutionary picture for low--luminosity AGN 
compared to the high--luminosity QSOs. Obviously,  
the rare, high--luminosity objects can form and 
feed very efficiently rather early in the 
universe. Their space density declines by
more than two orders of magnitude at redshifts below z=2.
The bulk of the AGN, however, has to wait much 
longer to grow with and shows a decline of space density
by less than a factor of 10 at redshifts below one. 
The late evolution 
of the low--luminosity Seyfert population is very 
similar to that which is required to fit the 
Mid--infrared source counts and background 
(Franceschini et al., \cite{fra02}) and 
also the bulk of the star formation in the 
Universe (Madau et al., \cite{mad98}), while the 
rapid evolution of powerful 
QSOs traces more the merging history of 
spheroid formation (Franceschini et al. \cite{fra99}).

This kind of anti--hierarchical Black Hole growth 
scenario is not predicted in most of the semi--analytic
models based on Cold Dark Matter structure formation
models (e.g. Kauffmann \& Haehnelt \cite{kau00}; 
Wyithe \& Loeb \cite{wyi03}). This could indicate two modes 
of accretion and black hole growth with 
radically different accretion efficiency 
(see e.g. Duschl \& Strittmatter \cite{dus02}). A
self--consistent model of the black hole growth which
can simultaneously explain the anti--hierarchical 
X-ray space density evolution and the local black 
hole mass function derived from the $M_{\rm BH}-\sigma$
relation assuming two radically different modes of
accretion has recently been presented by Merloni 
(\cite{mer04}).

 \begin{figure*}
 \centering
 \includegraphics[width=13cm]{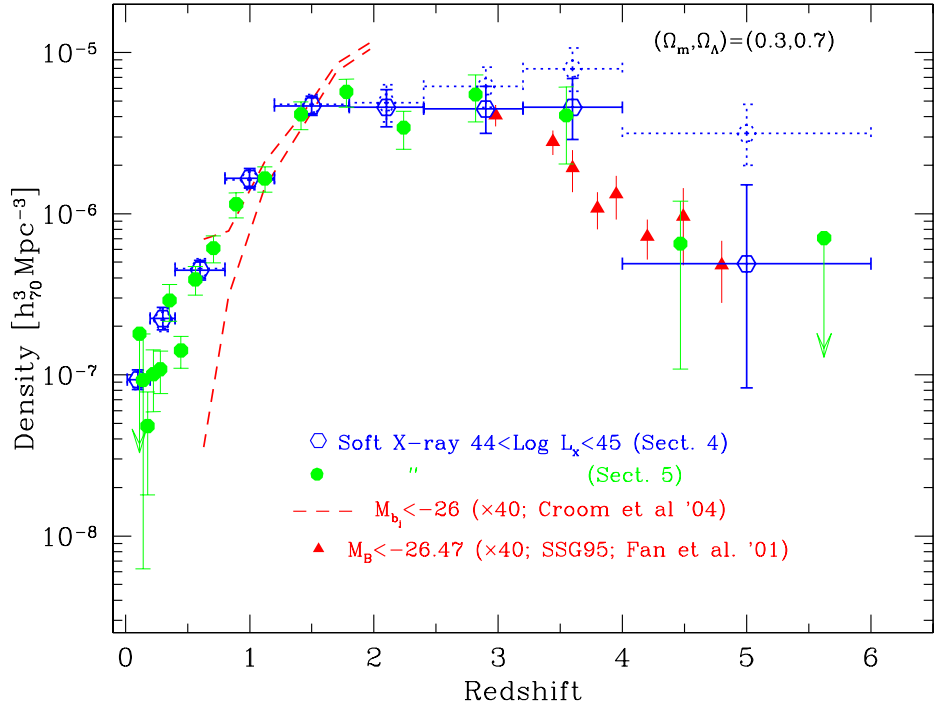}
  \caption{Comparison of the space density of luminous QSOs
    between optically selected and soft X--ray selected samples.
    The X--ray number densities are plotted for the luminosity 
    class $\log L_{\rm x}=44-45$, both for the 
    binned and $V_{\rm max}$ analysis with the same symbols 
    as in Fig.~\ref{fig:comp_ma_ta}.
    The dashed lines in represent the one sigma range for 
    $M_{b_{\rm j}}<-26.0$ from Croom et al \cite{croom04},
    multiplied by a factor of 40 to match the X--ray 
    space density at z=1.5. 
    The triangles at $z>2.7$ with  1$\sigma$ errors are from 
    Schmidt, Schneider, \& Gunn (1995) (SSG95) 
    and Fan et al. (2001) after a cosmology conversion (see text) 
    and a scaling by a factor of 40 to match with the soft X--ray
    density at $z\sim 2.7$. As discussed in this paper, both the rise and
    the decline of the space density change with $L_{\rm x}$ and 
    therefore this comparison can only be illustrative. 
   }
 \label{fig:zevcomp}
\end{figure*}

Finally, we compare the space density of soft X--ray selected 
QSOs from  
our sample to the one of optically-selected QSOs at
the most luminous end. The comparison is plotted in 
Fig.~\ref{fig:zevcomp}.  The $z<2$ number density curve for optically 
selected QSOs ($M_{b_{\rm J}}<-26.0$) is from the combination of the 
2dF and 6dF QSO redshift surveys by Croom et al. (\cite{croom04}). 
The $z>2.7$ number densities from Schmidt, Schneider \& Gunn (\cite{ssg95})
and Fan et al. (\cite{fan01}) have been originally given for 
$h_{70}=5/7, \Omega_{\rm m}=1, \Omega_{\rm \Lambda}=0$.
Their data points have been converted to our default cosmology
and the $M_{\rm B}$ threshold has been re-calculated with an
assumed spectral index of $\alpha_{\rm o}=-0.79$ 
($f_\nu\propto\nu^{\alpha_{\rm o}}$), following e.g. 
Vignali et al. (\cite{vig03}).
The plotted curve from Schmidt, Schneider \& Gunn / Fan et al. 
is for $M_{\rm B}<-26.47$ under these new
assumptions. A small correction of densities due to the cosmology 
conversion causing redshift-dependent luminosity thresholds 
has also been made, assuming 
$d\Phi/d\,{\rm log\,}L_{\rm B}\propto L_{\rm B}^{-1.6}$ 
(Fan et al. \cite{fan01}).  The space density for the soft X--ray QSOs 
for the luminosity class $44<{\rm log}\;L_{\rm x}<45$ has been 
plotted both for the binned and $V_{\rm max}$ determination. 
The Croom et al., (\cite{croom04}) space density  
had to be scaled up by a factor of 16 in order to match 
the X-ray density at $z\sim 2$. The Schmidt, Schneider \& Gunn / Fan et al. 
data points have been scaled by a factor of 40 to match the soft 
X--ray data at $z=2.7$ in the plot. There is relatively little 
difference in the density functions between the X--ray and optical
QSO samples, although we have to keep in mind, that both the rise 
and the decline of the space density is varying with X--ray luminosity,
so that this comparison can only be illustrative until larger 
samples of high--redshift X--ray selected QSOs will be available.  

Very recently, Wall et al. \cite{wal05}, have presented an update of the 
space density evolution of the Parkes quarter--Jansky sample of 
flat--spectrum radio sources. They basically confirm the rise 
and fall of the QSO population as now seen both in the optical and 
X-ray QSO populations. 

\section{Conclusions and outlook}

We have merged the 
{\em Chandra} and {\em XMM--Newton} deep survey data 
with the whole body of previously identified 
{\em ROSAT} AGN samples. We have selected only 
the type--1 AGN in all samples and treated only 
the detections and X--ray fluxes in the 0.5--2 
keV band. The different 
samples cover an unprecedented five 
orders of magnitude in flux limit and six 
orders of magnitude in survey solid angle 
between the {\em ROSAT} Bright and serendipitous surveys,
the {\em XMM--Newton} Lockman Hole survey   
and the {\em Chandra} Deep Surveys. 
The sample comprises 944 identified AGN--1 objects 
and only 57 unidentified sources, which could be  
AGN--1, i.e. roundabout 1000 objects. 
The luminosity--redshift diagram
is almost homegeneously filled with our sample
objects.
With this sample we arrive at the following
conclusions:

\begin{enumerate}

      \item 
      The new {\em Chandra} and {\em XMM--Newton} sources are 
      predominantly Seyfert galaxies at a median 
      luminosity of $\sim10^{43}$ erg s$^{-1}$ and a median 
      redshift around 0.7 and push the determination of the 
      X--ray luminosity and space density functions into
      so far unexplored parameter ranges of redshift 
      and luminosity. 

      \item AGN--1 are by far the largest contributors 
      to the soft X--ray selected samples. Their evolutional
      properties are responsible for the break in the total
      X--ray sources counts in the 0.5--2 keV band.

      \item The soft X--ray luminosity function of 
      AGN shows a clear change of shape as a function of 
      redshift, confirming earlier reports of 
      luminosity--dependent density evolution for 
      optical quasars and X--ray AGN.
 
      \item The space density function
      changes significantly for different luminosity classes.
      It shows a strong positive evolution, i.e. a density 
      increase at low redshifts up to a certain redshift and 
      then a flattening. The redshift, at which the evolution
      peaks, changes considerably with X--ray luminosity,
      from $z\approx0.5-0.7$ for luminosities 
      $\log L_{\rm x}=42-43$ erg s$^{-1}$ to $z\approx2$ for  
      $\log L_{\rm x}=45-46$ erg s$^{-1}$.

      \item The amount of density evolution from redshift zero 
      to the maximum space density is also a strong function 
      of X--ray luminosity. The change is more than a factor
      of 100 at high luminosities, similar to what has been
      observed for optically selected QSOs, but it is less
      than a factor of 10 for low X--ray luminosities.

      \item For the first time, we find  a clear decline
      of the space density of X--ray selected AGN towards
      high redshift, using a rigorous treatment of 
      optical incompleteness and the corresponding 
      survey volume. The decline is observed 
      clearly for X--ray luminosities in the 
      range  $\log L_{rm x}=42-45$~erg~s$^{-1}$, while at 
      higher luminosities the survey volume at high--redshift  
      is still too small to obtain meaningful densities.

   \end{enumerate}

In the future, X--ray surveys which are both wide and deep
are necessary, in order to provide enough volume for 
a better measurement of the space density function of the 
rare high--luminosity AGN at large redshifts. Several new 
surveys towards this goal are already underway, e.g. the 
{\em Chandra Multiwavelength Project (Champ)} (Silverman 
et al., \cite{silver04}), the {\em XMM-Newton AXIS project}
(Barcons et al., \cite{barc02}), the {\em Chandra Large Area Synoptic 
X-Ray Survey (CLASXS)} (Yang et al., \cite{yan04}; Steffen
et al., \cite{ste04}), the {\em Extended Chandra Deep 
Field South} (PI: W.N. Brandt) or the {\em XMM-Newton 
COSMOS Field} (PI: G. Hasinger), which together should enrich the sample 
of $z>4$ objects by about an order of magnitude. Ultimately,
new X--ray Dark Energy missions, aiming to survey large solid angles 
on the sky to considerable depth could provide a 
factor of 100--1000 increase in AGN sample size. 

\begin{acknowledgements}
      Part of this work was supported by the German
      \emph{Deut\-sches Zen\-trum f\"ur Luft- und Raumfahrt, DLR\/} grant
      number 50 OR 0207. Also this work is partially supported by
      NASA grant NAG 5-10875 (LTSA). TM appreciates support from
      Max-Planck Society during his visits to MPE. We thank an anonymous 
      referee for constructive comments, which significantly improved the 
      paper.
\end{acknowledgements}

\end{document}